\documentclass{aa}  
\def\SPFU{(2.6$\times$10$^{8}$) cm$^{-2}$ s$^{-1}$}

\def\deg{$^{\circ}$}

\newcommand{\nHat}[1]{\widehat{#1}}
\usepackage{graphicx}
\usepackage{subfigure}
\usepackage{txfonts}
\usepackage{longtable,lscape,rotating}
\usepackage{multirow}
\usepackage{array}
\usepackage{color}
\begin{document}

%
%

\title{OVII and OVIII line emission in the diffuse soft X-ray background: heliospheric and galactic contributions}

%
%
\author{D.Koutroumpa\inst{1} \and F.Acero\inst{2} \and R.Lallement\inst{1} 
\and J.Ballet\inst{2} \and V.Kharchenko\inst{3}}
%
\offprints{Dimitra Koutroumpa, \email{dimitra.koutroumpa@aerov.jussieu.fr}}
\institute{UMR 7620, IPSL/Service d'A\'eronomie, CNRS, 
Universit\'e Pierre et Marie Curie, Universit\'e Versailles-Saint-Quentin, Verri\`eres-le-Buisson, France
\and UMR 7158, DSM/DAPNIA/SAp, CEA Saclay, 91191 Gif-sur-Yvette, France
\and Harvard-Smithsonian Center for Astrophysics, Cambridge, MA, USA}
\date{Received 13/07/2007 / Accepted 10/09/2007}
%
%

\abstract{}
{We study the 0.57 keV (O VII triplet) and 0.65 keV (O VIII) diffuse emission generated by charge transfer collisions between 
solar wind (SW) oxygen ions and interstellar H and He neutral atoms in the inner Heliosphere. These lines which dominate 
the 0.3-1.0 keV energy interval are also produced by hot gas in the galactic halo (GH) 
and possibly the Local Interstellar Bubble (LB).}
{We developed a time-dependent model of the Solar Wind Charge-Exchange (SWCX) X-ray emission, based on the localization 
of the Solar Wind Parker spiral at each instant. We include input SW conditions affecting three selected fields, 
as well as shadowing targets observed with XMM-Newton, Chandra and Suzaku satellites and calculate X-ray emission 
in the oxygen lines O VII and O VIII in order to determine the SWCX contamination and the residual emission to attribute 
to the galactic soft X-ray background. We obtain ground level intensities and/or simulated lightcurves 
for each target and compare to X-ray data from the three instruments mentioned.} 
{The local 3/4 keV emission (due essentially to O VII and O VIII) detected in front of shadowing clouds is found to be 
entirely explained by the CX heliospheric emission. No emission from the LB is needed at these energies. 
The observed and modeled range of the foreground oxygen emission is 0.3-4.6 LU (Line Units = photons cm$^{-2}$ s$^{-1}$ sr$^{-1}$) 
for OVII and 0.02-2.1 LU for OVIII depending on directions and conditions.

Using the model predictions we subtract the heliospheric contribution to the measured emission and derive the halo contribution. 
We also correct for an error in the preliminary analysis of the Hubble Deep Field North (HDFN). 
We find intensities of 4.9$^{+1.29}_{-1.04}$ LU, 6.25$^{+0.63}_{-0.98}$ LU, 11.15$^{+2.36}_{-1.41}$ LU 
for OVII 
and 1.41$^{+0.60}_{-0.49}$ LU, 1.62$^{+0.35}_{-0.49}$ LU, 1.97$^{+1.11}_{-0.71}$ LU 
for OVIII towards the Marano Field, the Hubble Deep Field-North, and the Lockman Hole respectively.
}{}  
\titlerunning{Oxygen Line Emission in the Soft X-ray Background}
\authorrunning{KOUTROUMPA ET AL.}
\maketitle

\section{Introduction}
Charge-eXchange (CX) collisions between highly charged Solar Wind ions and solar system neutrals was identified as a very efficient 
mechanism of soft X-ray emission by \cite{cravens97}, following the discovery of X-ray emission from comets (\cite{lisse96}). 
Immediately after that, \cite{cox98} suggested that X-ray emission induced in Solar Wind Charge-eXchange (SWCX) 
with interstellar (IS) neutrals flowing across the Heliosphere would be an additional component of the diffuse 
soft X-ray background (SXRB), not to be considered lightly. 

Signs of the SWCX diffuse emission had already been detected during the ROSAT all-sky survey, as a background 
component varying in scales of one to several days and contaminating all observations with count rates comparable to 
the cosmic background (\cite{snowden93}). These temporal variations called long-term enhancements 
(LTEs; \cite{snowden95}) had been associated to solar wind variations (\cite{freyberg94}) although their origin remained unknown 
until the discovery of SWCX emission. 

\cite{cravens01} and \cite{robertson01} modeled the heliospheric SWCX emission with a simple model of the SW radial propagation 
without solar rotation, and explained satisfactorily the global correlation between the LTEs and strong solar wind enhancements. 
Moreover, the geocoronal emission was examined by \cite{robertson03a},b and measured during Chandra observations of the dark moon 
(\cite{wargelin04}). The temporal variations of the X-ray emission (including the ROSAT LTE's) were proved to 
be as much due to the geocoronal emission, as due to the heliospheric emission (H and He). However, the geocoronal emission 
consists mainly of short-term, intense peaks and is more tightly correlated to SW enhancements than the He and especially 
the H component of LTE's (fig.2 in \cite{cravens01}) which makes it much easier to identify. Their analysis showed that 
the geocoronal emission can reach the same order of magnitude as the heliospheric contribution during only short-time intervals. 

LTEs were, in general, easily identified in ROSAT lightcurves, and time-variable contaminated data were systematically removed 
(\cite{snow_frey93}), removing mainly the geocoronal component, but not all of the heliospheric emission. 
In general, it was proved that the contamination could be quite significant in the case of large 
solar wind enhancements, as has been demonstrated by the long-duration XMM-Newton exposure towards 
the Hubble Deep Field-North (Snowden et al., 2004, hereafter \cite{snowden04}). 

However, the ground level of the heliospheric SWCX emission has been a subject of debate. \cite{cravens00} estimated 
that it could be of the same order as the soft X-ray emission of the so-called Local 
Interstellar Bubble (hereafter LB). On the other hand, \cite{cox98} predicted that the heliospheric CX emission maxima 
should be aligned 
along the interstellar wind axis (i.e. towards galactic coordinates lII,bII= 4,+16° and in the opposite direction), 
a trend which was not observed in ROSAT survey maps. \cite{lall04} warned that parallax effects 
due to the ROSAT observation geometry destroy this axial symmetry and that as a consequence the absence 
of such maxima does not allow to draw conclusions. The reconstruction of the heliospheric emission sky map 
for the ROSAT survey geometry and the comparison with the data and the LB geometry suggested that, apart from a few 
specific regions and from high latitude directions, a significant contribution of the heliosphere is not precluded. 

The Local Interstellar Bubble is a $\leq $ 100 pc cavity surrounding the Solar System and presumably filled with rarefied, 
hot plasma. The idea of the Hot LB derived from the need to explain the fraction of the soft X-ray background 
that does NOT anticorrelate with the interstellar column density, and thus can not be attributed to distant galactic emission 
but needed a local origin (\cite{sanders77}; 
\cite{snowden90}). The temperature of the emitting gas has been inferred by the Wisconsin (\cite{fried80}; \cite{mccammon90}; 
\cite{snowden90}) and ROSAT (\cite{snowden93}) surveys band ratios in the 1/4 keV energy range, and found to be around 10$^6$ K.
 
Apart from distinct features like supernovae and superbubbles (e.g. Loop I), the diffuse SXRB below 1 keV is still largely 
considered to be consisting of three major components: (i) an unabsorbed 
$\sim$10$^6$ K thermal component originating from the LB, (ii) an absorbed $\sim$2$\times$10$^6$ K thermal 
component associated with the galactic halo and (iii) an absorbed extragalactic power law (\cite{kuntz00}, also 
see review of \cite{mccammon90} and references within).

In \cite{koutroumpa06} we have presented a thorough analysis of the total ground level SWCX heliospheric emission and exposed 
a detailed list of the factors influencing this emission, such as the solar cycle phase, the observer position, and the line 
of sight (LOS), as well as a preliminary analysis of a time-dependent model. 

In this paper we present a more detailed comparison 
between soft X-ray observations from XMM - Newton, Chandra and Suzaku and time-dependent simulations of heliospheric 
CX-induced soft X-ray emission. Our goal is to try and distinguish the post-CX heliospheric component within the soft X-ray 
background, in the range 0.5-1.0 keV and more specifically separate the oxygen (O VII 0.57 keV; O VIII 0.65 keV) 
heliospheric emission from the Local Interstellar Bubble (LB) and Galactic Halo emission in the same energy range. The two oxygen 
lines are of primary importance in the study of the LB, because if the 1/4 keV emission is usually analysed to derive the Bubble's 
temperature, the oxygen line ratio is used to constrain, or rule out, higher temperatures for the LB models (Henley et al. 2007a, 
here after \cite{henley07xmm}), as well as to constrain the halo emission characteristics.

X-ray data selection criteria and processing are presented in section \ref{data_proc}, followed by a thorough presentation 
of the time-dependent SWCX heliospheric model in section \ref{SWCXmodel}. Case to case results are detailed in sections \ref{mbm12} 
to \ref{lockman}. Finally, conclusions and global comparison to ROSAT 3/4 keV maps are discussed in section \ref{discussion}.

\section{Data selection and analysis}\label{data_proc}
\subsection{Selection criteria}\label{criteria}
The targets selected are summarized in table \ref{targets0} and can be assembled in two major groups 
according to the criteria upon which they were selected. 

\setlength{\extrarowheight}{1mm}
\begin{table}
\begin{center}
\caption[\small{\textit{Comparaison spectrale des modèles aux observations XMM}}]
{\footnotesize{List of selected targets.}\label{targets0}}
\small{
\begin{minipage}[t]{\linewidth}
\begin{tabular}{lccc}\hline \hline
Name & Gal. Coord. & Instrument & Obs. Period\\\hline

Marano Field& (269.8$^\circ$,-51.7$^\circ$) 
	& XMM & 22-30/08/2000\\ 

HDFN& (126.0$^\circ$,55.2$^\circ$) & XMM & 01/06/2001\\	 

Lockman Hole& (149.1$^\circ$,53.6$^\circ$) 
	& XMM & 15-27/10/2002\\ 
	
\multirow{2}{*}{MBM 12} & \multirow{2}{*}{(159.2$^\circ$,-35$^\circ$)} & Chandra & 17/08/2000\\
& & Suzaku & 03-08/02/2006\\

\multirow{2}{*}{Filament} & \multirow{2}{*}{(278.7$^\circ$,-46$^\circ$)} 
	& XMM & 03/05/2002\\
& & Suzaku & 01-03/03/2006 \\

\hline\hline 
\end{tabular}
\end{minipage}}
\end{center}
\end{table}

The initial goal of this study was to model short-scale variations of the soft X-ray heliospheric background due to temporary 
SW enhancements. This idea was inspired by the SWCX X-ray detection during the Hubble Deep Field North (HDFN here after) 
observation of June, 1, 2001 with XMM-Newton (\cite{snowden04}). The HDFN exposure was 
long enough to detect one of the most beautiful examples of CX emission in the Heliosphere, due to a large SW flux enhancement 
(\cite{snowden04}; Koutroumpa et al., 2006). 

Such long-lasting exposures are not very frequent in the observation schedules of X-ray observatories, so we decided to look in the 
XMM-Newton database for short-spaced repeated exposures on the same target-fields and preferentially associated with some medium 
or large SW enhancements recorded in the same periods by SW instruments (WIND, ACE/SWEPAM, OMNIWEB database). We thus collected 
three targets, including the HDFN, the Marano Field and the Lockman Hole, which compose our first group. We chose XMM-Newton 
because it disposes of a large Database of easy access. We needed a large spectral resolution in the 0.5-0.7 keV 
energy range, where the O VII and O VIII lines are dominant, so we used only MOS 1 and 2 spectra, which have the best resolution in this 
domain and the least calibration problems. The data extraction and analysis with XMM MOS 1 and 2 are 
detailed in \S \ref{sas} and \ref{xspec}.
 
Since the initial idea, the goal of this study has evolved. We decided to include two shadowing targets in the study, 
the molecular cloud MBM12 and a nearby filament in the Southern Galactic hemisphere (here after called South Galactic Filament = SGF). 
Each of these two shadows, has been observed during a minimum and maximum solar activity period. The observations are separated 
by several years, 2000 and 2006 for the MBM 12, and 2002 and 2006 for the SGF. Also, the observations were not made with the 
same instrument, Chandra and Suzaku for MBM 12, and XMM and Suzaku for the SGF. Even though the characteristics of these cases 
are quite different from the first group, we feel we can include them in this study since they are made in different solar 
conditions, and give an idea of the large-scale mean temporal variations of the SWCX X-ray emission level. Moreover, some of them 
are associated with quite particular SW conditions, either CMEs (MBM 12-Chandra observation, Filament-XMM observations), 
or sudden SW enhancement (MBM 12-Suzaku exposures). Each of the MBM 12 and SGF fields are detailed in paragraphs \ref{mbm12} 
and \ref{filament} respectively. 

In any case, all observation fields are as clear as possible of large diffuse sources, in order to distinguish the 
post-CX heliospheric emission. We avoid the galactic plane and compile target fields with as few bright point sources as possible. 

In table \ref{targets0} we list, besides the target name and central coordinates, the instrument used for each series 
of observations, and the observing period(s) for each field. Details on the data and model results will be given for each 
field in section \ref{results}. General discussions on the data-model comparisons will also be given in section 
\ref{discussion}.

\subsection{Data processing}\label{sas}
All XMM observations were reprocessed using the Science Analysis System (SAS version 6.5; the software is based on the model 
described in \cite{snowden04}). We used only MOS data to take advantage 
of its better spectral resolution. All the spectra used are an average of MOS1 and MOS2 detectors. To clean the event files 
from soft proton flare contamination, we used the MOS-FILTER process available in the Extended Source Analysis Software (XMM-ESAS). 
By building a histogram of counts/s in the 2.5-12 keV energy range and fitting a gaussian 
distribution upon it, it retains only the time intervals where the count rates are within a 3$\sigma$ range.

The different fields contain moderately bright sources that may have emission lines in the same energy bands as the charge exchange 
induced soft X-ray emission. The XMM SSC (Survey Science Center) produces a summary source list file providing information such 
as the coordinates and the flux of the detected sources in various energy bands for each observation. Using the 0.5-2 keV 
energy band, we removed all the sources that contribute to more than 1\%\ to the background emission in this particular energy band.
The region subtracted was a disk covering 90\%\ of the total flux of the sources. We then checked the results of the filter 
by building a "cheese" map (full field of view with sources removed) in a small energy band around the OVII and OVII 
emission lines (0.5-0.7 keV) to check that no sources remained. All spectra were extracted from this filtered event file. 
In this study we did not use any astrophysical nor instrumental background spectrum since the charge exchange induced 
X-ray emission is present in the entire field of view and any background spectrum may contain the emission we are trying to measure.

\subsection{Spectral modeling}\label{xspec}
The purpose of our simple model is to extract the flux of the O VII and  O VIII emission lines respectively at 0.57 keV and 0.65 keV. 
To do so with minimal assumptions, we add narrow lines at the expected energies to a continuum emission. In the imaging mode, 
the spectral resolution of XMM is much broader than any astrophysical line width. Concerning the continuum emission we assume, there 
are many origins to it but we decided to make a simple modelization with only two power laws.

We fit the soft X-ray spectra between 0.5 and 1.2 keV with XSPEC (11.3.2ad ; \cite{arnaud96}). The first component used in 
the model accounts for the residual soft proton contamination which is convolved by the redistribution matrix but not folded 
by the instrumental efficiency. The index and the normalization of this power law were free to vary.

The second component is an absorbed power law accounting for the extragalactic component. Its index was fixed to 1.46 
(\cite{chen97}) for all the observations. The absorption depends on the H column density of the part of the Galaxy each field 
is situated and the values are specified for each case in Sect.\ref{results}. We note that our modelling of the continuum 
is not realistic but it is sufficient for our study since we are mostly interested in charge exchange (CX) emission, 
which contains only lines.

Part of the spectral lines observed will be astrophysical emission from the Local Interstellar Bubble (LB) or a galactic corona. 
We prefer not to model this astrophysical emission to avoid underestimating the heliospheric component. We decided to model the 
following spectral lines with a gaussian distribution : O VII 0.57 keV, O VIII 0.65 and 0.81 keV, and Ne IX 0.91 keV.

The model (that we name PLC = PowerLaw Continuum) we used in our analysis can be described as follows: \textbf{powerlaw/b + 
phabs(powerlaw) + gaussian + gaussian + gaussian + gaussian}. The gaussian width was set to 0. This resulted most of the time 
in a good fit ($\chi^{2}\sim$ 140-170 for 135 d.o.f).

To test the robustness of our simple model for the continuum, we also tried a more sophisticated model for the continuum 
inspired by (and applied on) the SGF data analysis in \cite{henley07xmm} (See Sect. \ref{sgf_xmm}). This model consisted 
of a thermal plasma in collisional ionization equilibrium of temperature $T = 10^{6.06}$ K to model the LB, with oxygen 
abundance set to zero and two gaussians to account for the O VII and O VIII lines. The Galactic halo emission was modeled as 
two absorbed thermal plasma components of temperatures $T = 10^{5.93}$ K and $T = 10^{6.43}$ K respectively. 

In our test (that we name TCZEROX = Thermal Continuum ZERo OXygen), we used the LB model of \cite{henley07xmm} and 
we also forced a zero oxygen abundance for the halo components, keeping the same temperatures, so that all O VII and O VIII 
intensities are included in the gaussians. This test gave only a 15\%\ difference in the O VII strengths with respect 
to our simplified model, which gives us confidence in our procedure. The mean difference in the O VIII line flux is around 40\% , 
but does not affect our confidence, since this line measurements are much more uncertain than the O VII line. 
The results on the SGF data analysis are thoroughly detailed in section \ref{sgf_xmm}.

\section{Heliospheric SWCX Model}\label{SWCXmodel}
CX collisions, producing X-ray photons, are described by the reaction:
\begin{eqnarray}\label{eqn:yld}
{X^{Q+} \, + \,  [H, He]} & \rightarrow & {X^{*(Q-1)+} \, + \, [H^+, He^+]}  \\ 
 & \rightarrow & {X^{(Q-1)+} \, + \, [H^+, He^+] \, + \, [Y_{(h\nu ,H)}, Y_{(h\nu ,He)}]}\nonumber
\end{eqnarray}
where $Y_{(h\nu,H)}$, $Y_{(h\nu,He)}$ is the photon yield for the spectral line $h\nu$ induced in the CX between the ion $X^{Q+}$ with 
H and He respectively.

The model used to calculate the ground level heliospheric CX-induced soft X-rays is detailed in \cite{koutroumpa06}. It calculates 
the dynamical distribution of interstellar H ($n_H(r)$) and He ($n_{He}(r)$) atoms in the inner Heliosphere with respect to 
distance $r$ from the Sun, considering solar cycle variations effects and solar wind anisotropies for both H and He ionization processes. 

It then calculates the SW heavy ion radial propagation and loss due to 
CX occuring from collisions with IS atoms. We obtain then, the radial distribution of the SW heavy ion $N_{X^{Q+}}(r)$, 
depending on the density of ion $X^{Q+}$ at 1 AU:
\begin{equation}
N_{X^{Q+}o}=[X^{Q+}\,/\,O]\,[O\,/\,H^+]\,n_{H^+o}
\end{equation} 
and on an 
exponential term accounting for the ion loss due to CX with IS H and He, with cross sections $\sigma_{(H, X^{Q+})}$ 
and $\sigma_{(He, X^{Q+})}$ respectively. $[O]$ is the total oxygen ion content of the 
solar wind and $n_{H^+o}$ the proton density at 1 AU. The adopted values of $\sigma_{(H, X^{Q+})}$, $\sigma_{(He, X^{Q+})}$, 
$[X^{Q+}\,/\,O]$ and $[O\,/\,H^+]$ for the fast and slow solar winds are given in Table 1 of \cite{koutroumpa06}. 

For each selected target field and observation date, we calculate the path on the line of sight (LOS), 
decomposed in N$\sim$60 segments of increasing step $ds_j$ as we move away from the observer, from 0.1 AU at 1 AU up to 8.5 AU at 
the final distance of $\sim 85$AU, and 
the corresponding emissivity in units of (photons\, cm$^{-3}$ s$^{-1}$) :
\begin{equation}
\varepsilon _j\,  = 
R_{(X^{Q+},H)}(r)\,Y_{(h\nu ,H)} + R_{(X^{Q+},He)}(r)\,Y_{(h\nu ,He)} 
\end{equation}
depending on the spectral line $(h\nu )$ considered. 
\begin{equation}
R_{(X^{Q+},H)}(r) = N_{X^{Q+}o}(r)\,V_{SW}\,\sigma_{(H, X^{Q+})} \, n_H(r)
\end{equation}
and 
\begin{equation}
R_{(X^{Q+},He)}(r) = N_{X^{Q+}o}(r)\,V_{SW}\,\sigma_{(He, X^{Q+})} \, n_{He}(r) 
\end{equation}
is the volume collision frequency of ion $X^{Q+}$ with neutral 
heliospheric H and He respectively, in units of cm$^{-3}$ s$^{-1}$ and $V_{SW}$ is the SW mean speed, which approximates the 
relative speed between the SW ions and IS neutrals in the inner heliosphere, $\vec {v_{rel}} = \vec {V_{SW}} - \vec {v_n} $, 
since $v_n \ll V_{SW}$.

The integrated emission for the particular spectral line on the LOS is given by the following equation \ref{groundI}:
\begin{equation} \label{groundI}
\displaystyle I\, (LU) = (1\,/\, 4 \pi )\, \sum_{j=1}^{N} \varepsilon _j\, ds_j 
\end{equation} 
and defines the ground level emission of the spectral line for the particular date and LOS and the solar cycle phase 
(minimum or maximum) corresponding at this date. Details on the stationary model calculations are given in \cite{koutroumpa06}. 
All photon yields and cross-sections have also been calculated or discussed in previous papers 
(\cite{kharchenko00}; \cite{pepino04}). In 
the time-dependent model we use in this study we consider the triplet O VII at 0.57 keV, and the line O VIII at 0.65 keV only, 
since these lines are the best detected in the X-ray instruments we consider. In what follows, the time-dependent model described 
is equivalent for both lines considered.

Whenever solar instruments (WIND, ACE/SWEPAM) measure an important SW increase, due to solar flares, CMEs or other, within a few days 
of the observation dates of our selected fields, we apply a time-dependent simulation on the ground emission level 
to account for the variations induced in the total X-ray emission levels due to the enhancement. This was made for all of 
our targets except the Suzaku SGF exposure (Mars 2006) when the SW was rather calm. For each field we model the SW enhancements 
as one or multiple step functions, if there are more than one events during the observations period. The general procedure is similar 
for each simulation, based on the localization of the SW enhancements as a function of time along a Parker-type spiral, but 
we care to take into account the specific observation geometry and SW conditions for each target and date. These specific conditions 
will be detailed for each target in the sections \ref{mbm12} to \ref{lockman}. 

The general principle of modeling the impact of the SW enhancements on the segmented LOS is resumed in the following. 
Detailed calculations, formulas and schematic view of possible observation geometries are given in the appendix \ref{annexe}.
 
In our modeling we are compelled to consider two cases, whether the LOS is pointing forward on the Earth's orbit, 
or backwards on it, as explicitely shown in figure \ref{LOS_geo}. In fact, when the LOS is pointing forward on the orbit, 
the spiral is affecting the LOS progressively, starting from the observer on the Earth's position (fig. A.\ref{losfront}). 
In the second case of the LOS pointing backwards, the combined effect of the solar rotation and of the SW radial propagation 
acts in such a way that the LOS is affected starting at an intermediate point, dividing the LOS in two parts: 
(i) one on which the spiral is moving away from the observer, and (ii) a second (usually smaller) on which the spiral 
is approaching the observer (fig. A.\ref{losback}). We will go through the differences in modeling the two cases in the 
appendix \ref{annexe}.

The Active Region (AR) causing the SW enhancement is supposed to extend from North to South Solar Poles 
and is continuously emitting a SW proton ($p^+$) flux enhanced by a factor $f_{SW}(f_n)$ with respect to normal values 
$f_n$ = 2.6$\times$10$^8$ cm$^{-2}$s$^{-1}$, at a speed of $V_{SW}$. 
The solar longitudinal extend of the AR is defined such that the total duration of the enhancement is $\Delta t$ as measured in 
SW instruments. 

Equivalently, associated with the SW $p^+$ flux enhancement, we consider measured variation factor $[A]_{SW}([X^{Q+}\,/\,O])$ in 
the heavy ion relative abundances with respect to normal slow SW abundances $[X^{Q+}\,/\,O]$, in the particular circumstances 
the relative abundances of $O^{7+}$ (0.2) and $O^{8+}$ (0.07), obtained with ACE/SWICS intrument. 

At each instant $T_i$ we define the form of the Parker spiral taking into account solid solar rotation (27-day period), 
the radial propagation speed ($V_{SW}$), the 'ignition' time on the solar disk towards each radial direction ($T_{d_j}$), 
and the total width of the spiral ($\Delta D_{tot}$). Only in the cases of CMEs, we neglect the solar rotation, 
since radial propagation is dominating the CME structure.

Depending on the spiral's width and time $T_i$, the segment $ds_j$ can be affected either completely, or partially, as shown 
in the figures A.\ref{losfront} and A.\ref{losback}. This latter case is especially encountered when we model explosive CME's 
which are usually very brief. We include a correction $f_x$ and $A_x$, adjusting the enhancement factor and the abundance 
variation factor respectively, according to the total width of the sub-segment really touched by the enhancement 
at each instant $T_i$. These corrections are calculated for each LOS according to the observation geometry 
(LOS pointing forward or backwards, details in appendix \ref{annexe}).

Once we have defined the sub-segment(s) really affected by the spiral at instant $T_i$ we calculated the "new temporary" 
total heliospheric intensity $I_i$, modified because of the SW enhancement:
\begin{equation}\label{newI}
I_i (LU) = (1\,/\,4 \pi ) \,\, \sum_{j=1}^{N} \varepsilon ^{\prime}_j\, ds_j = (1\,/\,4 \pi ) \,\, \sum_{j=1}^{N} f_x\, [A]_x\, \varepsilon _j\, ds_j 
\end{equation}
We can then reproduce the temporal variation of the X-ray intensity levels during the periods of observation in simulated lightcurves 
for each of our targets. Abundance variations, can be correlated or anticorrelated with $p^+$ flux, so they can either 
emphasize or compensate for the influence of the SW $p^+$ flux enhancements, as we will detail in several examples.

The model is dealing only with soft X-ray emission generated in the Heliosphere, not accounting for 
the geocoronal emission. Although we cannot exclude that some residual geocoronal emission remains in the data and should be 
included in a more complete modeling of SWCX emission, we can consider, here, that the major part of the geocoronal emission is removed 
along with proton events (data processing, \S\ref{sas}), since it is exactly correlated with the SW variations (\cite{cravens01}). 
Moreover, a quick look on the observation geometries and the fact that XMM is in high orbit seems to indicate that most 
of the observations avoided the geocorona. Therefore, we decide to neglect any residual geocoronal emission in the SWCX analysis.

\section{Results}\label{results}
\subsection{MBM12 shadowing cloud}\label{mbm12}
MBM 12 is a nearby molecular cloud in the southern Galactic hemisphere (l, b = 159\deg .2, -34\deg ). It's distance is estimated 
with much uncertainty between 60 and 360 pc, according to various studies (\cite{hobbs86}; \cite{andersson02}; \cite{lall03}). MBM 12, 
with a column density of $N_H = 4\times 10^{21}$ cm$^{-2}$, is optically thick in the energy range 0.5 - 0.7 keV where the O VII 
(0.57 keV) and the O VIII (0.65 keV) lines are dominant. 
Therefore when observing on-cloud at this energy range, the emission flux detected must be generated nearby, partly by the CX 
process in the Heliosphere and a small residual emission due to the Galactic background. 

MBM 12 was frequently used as a shadow for the soft X-ray background, allowing to determine what fraction of 
the emission is generated in local regions, close to the Sun, and what fraction belongs to the Galactic (disk + halo) or 
extragalactic components. The first time the MBM 12 was observed as a shadow, was with ROSAT XRT/PSPC on July 31/ August 1 1991 
(Snowden et al., 1993, here after \cite{SMV93}), which yielded an upper limit for the observed ON-CLOUD emission of 
23 ROSAT Units and $\sim$ 75 ROSAT Units for the observed OFF-CLOUD emission (see fig. 3 in \cite{SMV93}).

ROSAT Units (RU) characteristic of the ROSAT measurements correspond to 10$^{-6}$ counts s$^{-1}$ arcmin$^{-2}$. An approximate 
equivalence between RU and LU (Line Units = photons cm$^{-2}$ s$^{-1}$ sr$^{-1}$) can be estimated based on \cite{smith07} 
reasoning on the line flux measured in ROSAT observations. \cite{SMV93} fitted the ROSAT data with a \cite{raymond77} model, 
simulating a standard LB of a temperature of 10$^6$ K which fits adequately the 1/4 keV band. 
For such a model, the 3/4 keV (0.5-0.8 keV) band produces only 47 counts s$^{-1}$ sr$^{-1}$ (= 3.98 RU, 
since 1 sr = 1.18$\times$10$^7$ arcmin$^2$) which are essentially due to the O VII triplet intensity producing 0.28 LU for such 
a model (according to ATOMDB). The PSPC instrument on ROSAT has very low resolution in the 3/4 keV band and cannot resolve 
O VII and O VIII lines of continuum. If we assume, keeping in mind the large uncertainties, that all of the emission detected 
in the 3/4 keV ROSAT band is due to the O VII triplet alone, then, 1 RU = 0.07 LU. With this equivalence we conclude that 
the intensity measured by ROSAT for the MBM 12 was $\sim$1.6 LU for the ON-CLOUD emission and $\sim$5.25 LU for the OFF-CLOUD emission.
 
\subsubsection{Chandra August 17, 2000}\label{mbm12_chan}
Chandra observed MBM 12 on August 17, 2000, for a total interval of $\sim$56 ks (\cite{smith05}). A first observation on July 9-10 
of the same year was excluded by the authors because of a severe solar flare. For the 17/08/2000 observation, the authors report 
an unexpected strong foreground emission in the O VII (0.57 keV) and especially the O VIII (0.65 keV) lines. Their best fit of the 
spectra yielded a foreground flux of $1.79\pm 0.55$ LU for the O VII line and $2.34\pm 0.36$ LU for the O VIII line. They discuss 
the possibility of the emission being due to a non-equilibrium nature of the LB, but rather conclude that their observations 
were contaminated by CX induced emission inside the Heliosphere.

\begin{figure}
\centering
\subfigure[Chandra-MBM 12]
{
\label{mbm12_cgeo}
\includegraphics[width=0.4\textwidth ]{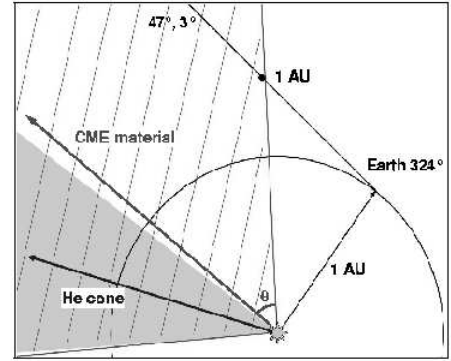}
}\\
\subfigure[Suzaku-MBM 12]
{
\label{mbm12_sgeo}
\includegraphics[width=0.4\textwidth]{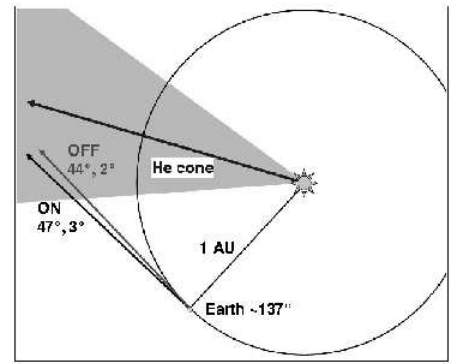}
}
\caption[\small{\textit{}}]
{\footnotesize{Geometries of the MBM 12 observations seen from the North ecliptic pole. 
\textit{Panel a}: Chandra-MBM 12 observation geometry on 17/08/2000 with the CME-1 material expanding in the interplanetary space. 
\textit{Panel b}: Suzaku-MBM 12 observation geometry for the period 03-08/02/2006 for both ON (black LOS) and OFF (red LOS) exposures. 
The He cone is crossed by both LOS, but the OFF exposure sees through denser He distribution. See details in text.}}
\label{mbm12_geoS}
\end{figure}

Indeed, on August 12, 2000 at 10:35UT, SOHO/LASCO observed a CME, which we name CME-1 here after, associated with an M-class flare 
from active region (AR) 9119, behind the west limb (\cite{edgar06}). The CME was on the high end of CME masses, 
with a total mass of $M = 1.2\times  10^{16}$g ejected, during a total of $\Delta t = 3$h. It's speed was evaluated at about 
$660$km/s. Apparently, the Chandra observation geometry (fig.\ref{mbm12_cgeo}) shows that Chandra was pointing through 
the interplanetary region affected by the CME material. A second CME (named CME-2) of equivalent mass, 
with a mean speed of $900$km/s and originating from AR 9114 near the first one, was recorded four hours later. In addition, 
ACE/SWICS measurements indicate that SW $O^{8+}$ had a relative abundance of 25\%  during the week preceeding the Chandra observation, 
and during the observation. 

We model the CME(s) propagation through the interplanetary space and how it affected the Chandra observation of MBM 12. In figure 
\ref{lasc3} (Online Material, here after OM)
we present a series of LASCO/C3 images of CME-1, begining at 11:18UT and ending at 14:18UT. In the figures, 
the inner white circle represents the Solar disk while the outer circle traces the sphere limits at 
$R\, =\, 6.4\, Solar\, Radii\, (R_\odot)$. This is the surface crossed by the CME in the temporal interval showed in the figures. 
We can consider, then, that all of the CME mass crossed this sphere in $\Delta t = 3$h and we can calculate the 
SW proton flux $f_{SW}$ in the CME, according to equation \ref{pflux_CME}:
\begin{eqnarray}\label{pflux_CME}
f_{SW} & = & M\,/\,(S\, \Delta t)
\end{eqnarray}  
where $S = 2\, \pi\, R^{\, 2}\, (1 - cos\theta )$ and $\theta$ is half the angular width of the CME. In a 3h interval, the Sun's 
rotation can be neglected with respect to the propagation speed of the CME material. Thus, from equation \ref{dep_time}, we define 
as departure time $T_{d_j}$ = 12/08/00 10:35UT the onset time of CME-1, as recorded on the solar disk in EIT UV and LASCO images. 
In the same way we can evaluate the CME-2 SW flux, and departure time as $T_{d_j}$ = 12/08/00 14:54UT.

The CME speed and angular width are the two parameters that influence the most the post-CX X-ray emission, because 
they define the part of the LOS to be affected and also the timing for the CME to impact on it. The LOS emissivity is maximum 
near the observer, up to 3 - 4 AU, therefore, from figure \ref{mbm12_cgeo} we understand that the CME width and speed 
will determine if the MBM 12 LOS is to be strongly affected during the Chandra observation or not.

The angular width of a CME is very difficult to evaluate since the apparent width can be much larger than the actual width 
because of projection effects, so that it reaches 360\deg\ for a CME originating at the 
Sun Center. CME-1 and CME-2 were classified as Partial Halo events because their total angular widths were estimated 
at $2\theta = 168$\deg\ and 161\deg\ respectively. The LASCO/C2 and C3 images of CME-1 (OM-fig.\ref{lasc3}) show a 
latitudinal apparent width of $\sim 40$\deg\ , but since the 
CME-1 occured just behind the west limb, it must have had a rather large width. CME-2 could have a smaller width as it was on the 
visible part of the Solar Disk. If we consider an intermediate angular width of CME-1 
and CME-2, $2\theta = 100$\deg\ and $2\theta = 60$\deg\ , we obtain from equation \ref{pflux_CME} a SW proton flux of 
$\sim$ 5$\times$ \SPFU and $\sim$ 10$\times$ \SPFU for each respectively. 

\begin{figure}
\centering
\includegraphics[width=0.5\textwidth ,height=!]{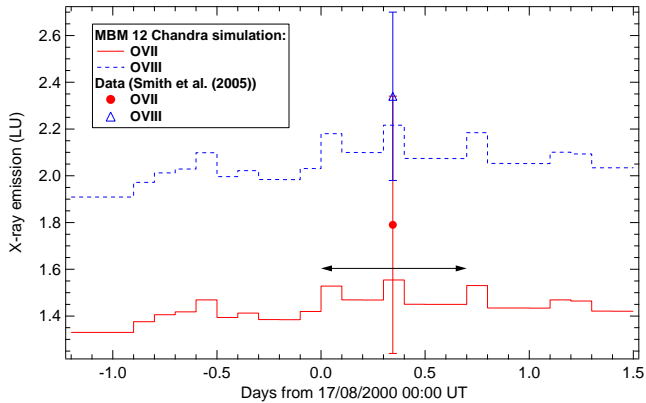}
\caption{\footnotesize{Simulated lightcurves of O VII (red plain line) and O VIII (blue dashed line) emission 
in LU for the MBM 12 Chandra exposure (17/08/2000 01:04UT) under the combined impact of CME-1 and CME-2 (see details in text). 
X-ray data is also presented (\cite{smith05}): O VII red dot, O VIII blue triangle. The horizontal double arrow 
delimits the Chandra exposure duration.}}
\label{mbm12_Clcurve}
\end{figure}

In this context, the LOS is 'touched' at a point near 1 AU away from the observer by CME-1 and around 2 AU away from 
the observer by CME-2 a little later. The simulated lightcurves of O VII and O VIII emission modified 
because of the interaction with CME-1 and CME-2 material are presented in figure \ref{mbm12_Clcurve}. This interaction, 
in combination with the strongly enhanced $O^{8+}$ relative abundance $\left[O^{8+}\,/\,O \right] = 0.25$, gave a mean photon flux of 
1.49 LU for the O VII triplet and 2.13 LU for the O VIII line in the same interval as the Chandra exposure. The ground level 
emission of the Heliosphere for the MBM 12 direction observed on the 17/08/2000 was modeled as a maximum solar activity period 
and yielded 1.33 LU and 0.52 LU respectively for O VII and O VIII lines, which means a 12\% increase for the O VII triplet 
and over 300\% increase for the O VIII line due to the abundance/CME perturbations. 

These values are very close to the ones measured by Chandra On-Cloud, which reinforces the conclusion of Smith et al. (2005) 
that the emission detected was entirely due to SWCX, not only for the O VIII line which was the most affected by the CMEs 
and the changes in abundances they induced, but also for the O VII triplet.  

\subsubsection{Suzaku February 3-8, 2006}\label{mbm12_suz}
The observation of MBM 12 with Suzaku on  February 3-8 2006 was performed in two consecutive exposures, ON-CLOUD (3-6/02/2006, 
for a total of 231 ks) and immediately after that, OFF-CLOUD (6-8/02/2006, for a total of 168 ks) (Smith et al., 2007). 
The ON-CLOUD exposure was pointing at galactic coordinates (159.2$^{\circ}$, -34.47$^{\circ}$), which translates to 47$^{\circ}$, 3$^{\circ}$ 
of helioecliptic coordinates, while the OFF-CLOUD exposure was about 3$^{\circ}$ away from the cloud, 
at galactic coordinates (157.3$^{\circ}$, -36.8$^{\circ}$) equal to 44$^{\circ}$, 2$^{\circ}$ in helioecliptic coordinates. 

ON-CLOUD the authors detect a local O VII line flux of (3.34 $\pm$ 0.26) LU and an O VIII line flux of (0.24 $\pm$ 0.10) LU, while 
OFF-CLOUD the total line flux, including local (SWCX + LB?) and unabsorbed distant (galactic disk + halo) emission, rises at 
(5.68 $\pm$ 0.59) LU and (1.01 $\pm$ 0.26) LU for O VII and O VIII respectively. The authors note that SW conditions were quite stable 
during the Suzaku exposures, except a short period of the ON-CLOUD observation, and conclude that the discrepancies with their previous 
results on MBM-12 observation with Chandra (\ref{mbm12_chan}; \cite{smith05}) were due to uncertainties on the background, 
especially in the case of Chandra, as well as to the large solar flare/CME influencing the Chandra background even more. 
However, they acknowledge 
once more the probability that SWCX emission may be contaminating at an unknown degree the MBM 12 O VII and O VIII detection.

In this study, we model the MBM 12 Suzaku observations, taking into account the short SW perturbation recorded at the end 
of the ON-CLOUD pointing and the influence it had on the measured data. We note first three factors 
partly responsible for: (i) the Chandra - Suzaku discrepancy and (ii) the ON/OFF difference during the Suzaku observations. 

The first point is the very strong difference between the solar maximum conditions in 2000 (Chandra observations) 
and the 2006 solar minimum conditions (Suzaku data). Indeed, 
as demonstrated in the monochromatic maps in \cite{koutroumpa06}, for low helioecliptic latitudes $\pm 20^{\circ}$ the SWCX X-ray 
emission is higher for solar minimum than for solar maximum. This is due to the fact that at solar minimum the radiation pressure 
is weakened with respect to gravity and the IS H and He trajectories are more convergent. This, in addition with the fact that ionization 
processes are less efficient, favours a better concentration of IS neutrals around the Sun (and the observer), filling more efficiently 
the ionization cavity for H, and enlarging the gravitational cone extent and density for He atoms. Therefore, it is quite logical that 
Suzaku measurements find a higher O VII and O VIII levels, since the SWCX emission is closer and brighter at solar minimum.

Second, we must point out the observation geometry differences between the Chandra and Suzaku exposures (fig. \ref{mbm12_geoS}). 
The Chandra observation, on August 17, 2000 was pointing downwind, but almost parallel to the He gravitational cone, thus missing 
the enhancement due to this part of the Heliosphere (Koutroumpa et al., 2006). On the contrary, Suzaku in February was pointing directly 
inside the He cone, which was only 1 - 2 AU away from the observer, where the emissivity on the LOS is maximum. Therefore, 
the Suzaku measurements should be normally much larger than the Chandra data. 

Finally, the OFF-CLOUD LOS of Suzaku was pointing through a denser region of the He cone with respect 
to the ON-CLOUD LOS in the same period. Consequently, the OFF-CLOUD emission should be slightly higher than the ON-CLOUD emission, 
even for calm SW conditions. Indeed, ground level O VII and O VIII fluxes for the ON-CLOUD pointing are 3.82 LU 
and 1.48 LU respectively, while for the OFF-CLOUD exposure they are 4.05 LU and 1.57 LU respectively. The ON-CLOUD ground level fluxes 
are $\sim$ 3 times higher than the equivalent fluxes in the Chandra simulation.

\begin{figure}
\centering
\includegraphics[width=0.5\textwidth ,height=!]{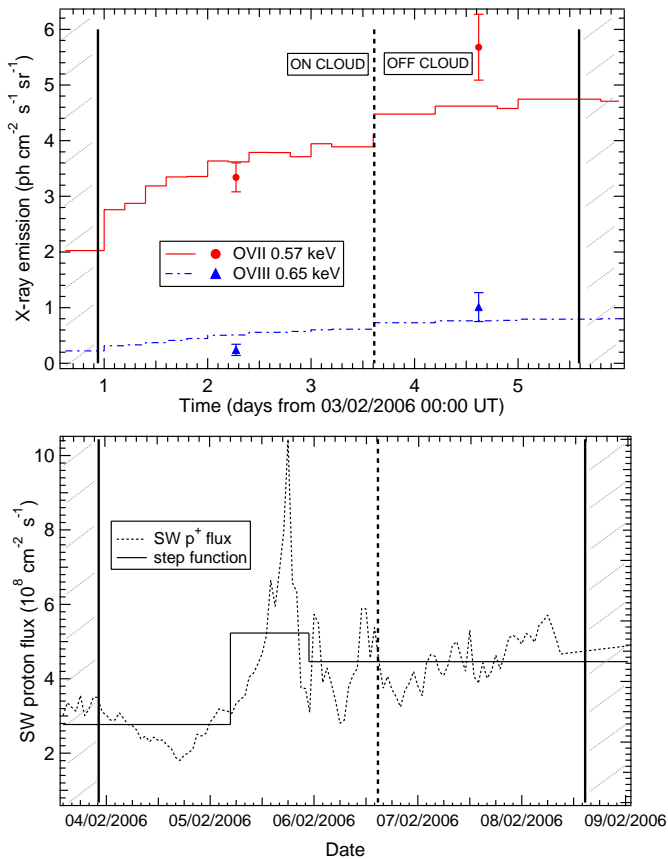}
\caption{\footnotesize{\textit{Upper Panel}: Simulated lightcurves for the O VII and O VIII line emission in Line Units 
(photons cm$^{-2}$ s$^{-1}$ sr$^{-1}$) for the MBM12 ON and OFF observations during the period 03-08/02/2006. 
The plain red line is for the O VII emission and the dashed blue line is for the O VIII emission. 
Red circles represent the measured ON and OFF O VII line flux, while the blue triangles represent 
the measured ON and OFF fluxes for O VIII line (Smith et al., 2007). 
\textit{Lower Panel}: Solar wind proton flux (dotted line) in units of $10^8$ cm$^{-2}$ s$^{-1}$ for the same period. 
The step function simulating the SW enhancement is presented with the plain black line.
In both panels, the vertical plain lines represent the start and end of the observation period, 
while the dashed vertical line is the separation between the ON and OFF exposures.}}
\label{mbm12suzaku}
\end{figure}

Naturally, only the difference of the observing geometry between ON and OFF cloud pointings is not enough to explain the ON/OFF difference 
in X-ray emission intensities. We model the temporal variations of the X-ray emission in the 0.5 - 0.7 keV including a step function 
of 2.01$\times$\SPFU, to account for the SW 'spike' recorded on day $T_o$ = 2.2, starting on 03/02/2006 00:00 UT, and propagating at a 
particularly low speed ($V_{SW}$ = 350 km/s) for 0.75 d. The SW proton flux remained at a high, but stable, level equal 
to 1.7$\times$\SPFU after the spike. The modeled step function is presented with the black plain line in the lower panel of 
figure \ref{mbm12suzaku}. 

$O^{7+}$ relative abundance was at about half its normal value ($\left[ O^{7+}\,/\,O\right]$ = 0.11) before the SW spike which 
reduced the ground level of around the same factor. During and after the SW enhancement the relative abundance remained at lower 
($\left[ O^{7+}\,/\,O\right]$ = 0.18) than normal values. Data for the $O^{8+}$ relative abundance are very poor, but indicate 
that $O^{8+}$ was also very scarce in the SW ($\left[ O^{8+}\,/\,O\right]$ = 0.01) before the SW enhancement, which also reduced 
the initial ground level. Abundance remains very low throughout all the Suzaku observations, so we considered a very slow recovery, with 
a value of 0.02 during the SW enhancement and 0.03 after that. 

The simulation results combining the difference in the ON and OFF observation geometries, the SW enhancement and reduced abundances 
are shown in the upper panel of figure \ref{mbm12suzaku}.
The X-ray emission starts at a lower than ground level, due to the initial very low abundances of both $O^{7+}$ and $O^{8+}$. Then, 
the model is predicting a rise in both ON and OFF X-ray background emission, which accentuates even more the 
difference due to geometry effects. The mean O VII line fluxes for the ON and OFF observations are 3.56 LU and 4.62 LU  respectively. 
On the other hand, the O VIII line fluxes for the ON and OFF exposures are 0.5 LU and 0.77 LU respectively. The ON-CLOUD simulation 
values for both O VII and O VIII lines are slightly above, but within error bars, the corresponding measured data, strongly implying 
that the foreground local emission detected on the MBM 12 cloud could be exclusively originating from within the Heliosphere. 

On the other hand, for the OFF-CLOUD pointing the model results are lower than the measured values. Nevertheless, it is strongly 
implied that besides the large ground level foreground contamination from heliospheric X-rays, the distant component could be lower 
than what was suggested by \cite{smith07}, due to local temporal enhancements of the heliospheric component. Indeed, the $\sim$30\% 
and $\sim$55\% increase respectively in the O VII and O VIII line intensities have been attributed to the Galactic Halo 
oxygen emission, since in the \cite{smith07} analysis the foreground emission is considered stable. In this case, what was 
attributed to distant galactic disk and halo should be revised to the residual flux of O VII and O VIII lines noted 
in table \ref{shadows}.

\subsection{South Galactic absorbing Filament}\label{filament}
\cite{henley07xmm} and \cite{henley07suz} have observed an absorbing filament in the Southern galactic Hemisphere. They fitted 
simultaneously the ON-Filament and OFF-Filament exposures obtained with XMM-Newton (\cite{henley07xmm}) 
and Suzaku (Henley et al., 2007b) respectively. 
Their goal was to use the difference in the galactic column density in the two directions (ON: 9.6$\times$10$^{20}$ cm$^{-2}$ 
and OFF: 1.9$\times$10$^{20}$ cm$^{-2}$) in order to disentangle the unabsorbed foreground emission attributed to the LB 
and the absorbed extragalactic and galactic halo component. 

Their basic spectral modeling was the same for both XMM and Suzaku observations and was composed of: (i) a thermal plasma model 
in collisional ionization equilibrium for the LB, (ii) two absorbed thermal plasma components for the Galactic halo, 
and (iii) an absorbed power law for the unresolved extragalactic sources (AGN). The absorption was determined by the column density 
for the ON and OFF exposures as mentioned before.

In the following paragraphs \ref{sgf_xmm} and \ref{sgf_suz} we will detail the results obtained  by \cite{henley07xmm} 
and \cite{henley07suz} for the O VII and O VIII line emission in the case of XMM and Suzaku respectively, as well as 
the SWCX simulations we applied on the XMM observations.

\subsubsection{XMM exposures 03/05/2002}\label{sgf_xmm}
XMM observed the SGF on May 3, 2002 in two consecutive exposures ON and OFF for a total 12.8 ks and 27.8 ks respectively. 
After flare removal with standard XMM data analysis procedures, the useful time remaining was 11.9 ks for the ON exposure and 
only 4.4 ks for the OFF exposure.

\cite{henley07xmm} fitted the ON and OFF spectra obtained with MOS-1 and 2 cameras simultaneously, applying the standard model described 
previously. Their standard LB model assumed a thermal plasma with a temperature of 10$^{6.06}$ K and emission measure 
E.M. = 0.018 cm$^{-6}$ pc. For these parameters, the LB standard 
model line emission yields 2.9 LU for O VII and 0.017 LU for O VIII (\cite{henley07xmm}, ATOMDB database).  

In order to measure more precisely the O VII and O VIII line emission from the LB in the XMM data, they replaced the 
standard thermal plasma model with a variation including two Gaussians for the oxygen lines and a thermal plasma model with 
frozen parameter of zero oxygen abundance to account for the continuum and remaining spectral lines. The Gaussians' widths were fixed 
at zero and the energies were fixed at 0.5681 keV for the O VII triplet and 0.6536 keV for the O VIII line. The two Gaussians 
yield O VII and O VIII line fluxes of 3.4$^{+0.6}_{-0.4}$ LU and 1.0 LU respectively, which is higher than what is predicted 
by the standard LB models (\cite{henley07xmm}, ATOMDB database).

\setlength{\extrarowheight}{2mm}
\begin{table*}
\centering
\caption[\small{\textit{}}]
{\footnotesize{Summary of O VII and O VIII modeled and measured line fluxes for the XMM/SGF May, 3 2002 observation. 
\textit{NB}: O VII and O VIII line fluxes in each column do not refer to the same diffuse SXRB component. 
SWCX model refers to the heliospheric component, \cite{henley07xmm} results to the line intensity attributed to the LB, 
and XMM/PLC and XMM/TZEROX to the total intensity measured in the O VII and O VIII gaussians in the present XMM data analysis. 
For details see text in \S \ref{sgf_xmm}.}\label{SGF_table}}
\begin{minipage}[t]{\linewidth}
\begin{tabular}{cccccccc}\hline \hline
&\multicolumn{2}{c}{\underline{\,\, SWCX\,\, }} & \underline{\,\, \cite{henley07xmm}\,\, } 
	& \multicolumn{2}{c}{\underline{\,\, XMM/PLC\,\, }} & \multicolumn{2}{c}{\underline{\,\, XMM/TCZEROX \,\, }} \\
& ON & OFF & LB & ON & OFF & ON & OFF \\\hline 
O VII & 3.16  & 3.47 & 3.4$^{+0.6}_{-0.4}$ & 11.38$^{+1.51}_{-1.65}$ & 16.95$^{+2.66}_{-2.67}$ & 13.23$\pm1.37$ & 16.55$\pm2.5$ \\
O VIII & 1.02 & 1.11 & 1.0 & 3.36$^{+0.73}_{-0.70}$ & 2.74$^{+1.10}_{-1.06}$ & 5.07$^{+0.68}_{-0.64}$ & 3.52$^{+1.10}_{-1.06}$ \\\hline \hline
\end{tabular}
\end{minipage}
\end{table*}

We need, at this point, to comment on the central energy position of the O VII triplet. The O VII central energy was biased 
by the assumption that emission is due to thermal plasma, and was chosen as the mean energy of the resonance (O6r : 574 eV), 
intercombination (O6i : 568.5 eV) and forbidden (O6f : 560.9 eV) lines, weighted by the line emissivities 
for a 10$^{6.06}$ K plasma. Indeed, the location of the centroid of the O VII triplet depends on the emission mechanism considered 
and can be represented by the line ratio $G = (O6f +O6i)\,/\,O6r$. In the case of hot plasmas ratio $G$ is less than 1, 
but larger than 3 in the case of charge-exchange (\cite{kharchenko05}). Future instruments with better spectral resolution should 
be able to separate the oxygen triplet lines and give more information on the nature of the soft X-ray background. 

The XMM-Earth system was positioned at a heliospheric longitude of 222\deg\ on  03/05/2002, and the SGF LOS were pointing at 
high south heliospheric latitudes (ON: 352\deg , -75\deg , OFF: 353\deg , -73\deg ). The SWCX model for the SGF observation 
geometry on 03/05/2002, considered with maximum solar conditions, predicts a ground 
level O VII line emission of 2.32 LU and an equivalent O VIII line emission of 0.92 LU. Near-Earth SW measurements show moderate 
variations, but not neat enhancements to be modeled as step functions. 

However, in the LASCO/CME catalog we find a CME starting on 30/04/2002 23:32UT ($T_o + \Delta t$ = -2.02 + 0.167 d), 
and progressing at a speed of $V_{SW}$ = 1100 km/s centered at a helioecliptic longitude of about 307$^\circ$ and 
with half width angle $\theta \sim$ 65\deg . These parameters give a SW proton flux at $f_{SW}$ = 6$\times$\SPFU . If we adopt 
moderately enhanced relative abundances during the observation period for both $O^{7+}$ and $O^{8+}$, 0.30 and 0.08 respectively, 
we obtain the following line fluxes  for the ON and OFF exposures: 3.16 LU (ON) and 3.47 LU (OFF) for the O VII triplet 
and 1.02 LU (ON) and 1.11 LU (OFF) for the O VIII line. These simulated flux values, along with those obtained by \cite{henley07xmm} 
and attributed to the LB, as well as those measured for the total line intensity (foreground + distant halo component) in the 
two XMM data fits (PLC, TCZEROX) are summarized in table \ref{SGF_table}.

Compared to \cite{henley07xmm} results on the LB line intensities, both ground level and especially CME-modified line intensities 
could account for most (if not all) of the emission attributed to the LB (3.4 LU and 1. LU respectively for O VII and O VIII). 
Moreover, the CME impact on the OFF-SGF exposure, produces a 10\% enhancement with respect to the ON-SGF exposure, especially in 
the O VII intensity. \cite{henley07xmm} assumed that the foreground LB emission was identical for the two exposures, thus, the 
10\% enhancement of the heliospheric component was probably erroneously attributed to the Galactic Halo. We regret that 
\cite{henley07xmm} did not provide equivalent line intensity values for the galactic halo to compare to. 

In our XMM data processing and modeling we made no assumptions on the nature of the O VII and O VIII line emission 
(see \S \ref{xspec}), and removed only the continuum from XMM data (columns XMM/PLC and XMM/TCZEROX in table \ref{SGF_table}). 
Subtraction of the SWCX heliospheric component should give the residual emission, to really be 
separated in LB and Galactic halo emission, which can radically modify the parameters initially assumed for those components. 

SGF is a medium-size absorber. In fact, \cite{henley07xmm} comment the fact that the SGF-ON column density is much lower than 
other shadowing observations: $4\times 10^{21}$ cm$^{-2}$ for MBM 12 (Smith et al., 2005, 2007), up to $\sim 10^{23}$ cm$^{-2}$ 
for Barnard 68 (\cite{freyberg04}, which gave higher temperature results for the LB (10$^{6.21}$ K). For these absorbers, 
it is easier to conclude that what is measured as foreground emission is really local (LB according to \cite{henley07xmm}, 
SWCX according to our study), because transmissivity of O VIII radiation is less than 14\% . On the contrary, the SGF has a 
higher transmissivity (up to 61\% for O VIII energies) and therefore it is more difficult to really determine what originates 
from the local component or leaks through the filament from distant components. 

In their study, \cite{henley07xmm} test the case of a hotter LB (10$^{6.21}$ K), which would give higher O VIII emission, but 
conclude that such a case predicts far too much intrinsic halo O VI intensity, which is incompatible with FUSE 
measurements (\cite{shelton07}). They are, thus, constrained to admit that the different temperatures of the LB resulting 
from the various shadowing observations, are due to anisotropies inside the LB for the different LOS. 

What we could propose is the contrary: if the emission attributed to the LB is largely or entirely heliospheric, as it is suggested in our analysis, 
there is no need to look for a hotter or anisotropic bubble and hence, no need for a cooler halo and no discrepancy with FUSE measurements.

\subsubsection{Suzaku exposures 01/03/2006}\label{sgf_suz}
The Suzaku observations of the SGF were performed on March 1, 2006 16:56 UT for a total of $\sim$80 ks (ON-Filament), and 
on March 3, 2006 20:52 UT for a total of $\sim$100 ks (OFF). \cite{henley07suz} fit the same basic model 
(unabsorbed LB + absorbed$\times$(halo + extragalactic)) on the spectra and find LB and halo parameters very different from those obtained 
with the XMM data in 2002. In detail, they detect a much fainter O VII triplet of 0.17 LU only for the LB, while O VIII line flux is 
not detected with certainty. No details on the halo absolute line fluxes for the two oxygen lines are given. 

The authors are perplexed by the difference of their Suzaku spectra of the SGF and their equivalent XMM spectra 
previously analyzed in \cite{henley07xmm}. They report steady SW conditions within normal measured values, and claim that "the SWCX 
emission should be the same for both (XMM and Suzaku) observations, composing a fixed fraction of the foreground emission".

At this point, we need to stress once more the effect of the solar cycle phase on the SWCX component and thus on 
the soft X-ray data. As we mentioned earlier, there is a large difference on the O VII and O VIII line fluxes 
between solar maximum (years 2000-2002) and solar minimum (2006), because of the difference in the distributions of IS neutrals 
in the Heliosphere. In the case of SGF, which points at very high southern ecliptic latitude (353$^\circ$,-73$^\circ$) the effect is 
the opposite with respect to MBM 12, which was in low ecliptic latitude, because it is dominated by differences in the solar ions 
relative abundances. 

At solar maximum, the SW is merely isotropic and the ion relative abundances are homogeneous and close to slow SW conditions. On the 
other hand, at solar minimum we find a SW highly anisotropic and composed of equatorial regions of slow wind and high latitude regions 
of fast wind. In this latter case, LOS pointing at high ecliptic latitudes are partly affected by fast SW (above $\sim 20^\circ$) 
and the X-ray emission is dominated by differences of ion relative abundances with respect to maximum cycle phase. Indeed, 
$O^{8+}$ is completely absent from the fast SW and $O^{7+}$ is strongly depleted ($\left[ O^{7+}\,/\,O\right]$ = 0.07). 
Therefore, X-ray emission on high latitude LOS is expected to be much fainter at solar minimum than solar maximum.

During the Suzaku observations of the SGF SW flux was very calm, but also there was rather low $O^{7+}$ and $O^{8+}$ 
relative abundances as recorded by ACE/SWICS in the ecliptic plane (in general the ecliptic plane is dominated by 
slow SW conditions). The $O^{7+}$ was only 0.083 during the ON-Filament exposure, while it rose to 0.13, 1.5 d after, 
for the OFF exposure. $O^{8+}$ statistics were very poor in the SWICS data for this period, but show very low values, 
of 0.02 for the period of the exposures. For lack of data, we assume that high latitude abundances were lower 
than average by the same amount as in the ecliptic.

The ground level model predicts an O VII line flux of 0.83 LU and an O VIII line flux of 0.07 LU (for solar minimum 
conditions in the LOS of the SGF). After abundance correction (0.083/0.2 for $O^{7+}$ and 0.02/0.07 for $O^{8+}$) 
we predict 0.34 LU in O VII and 0.02 LU in O VIII for the ON-Filament exposure. These values are about an order of magnitude 
below the equivalent values in the SGF-XMM simulation, which is the same difference \cite{henley07suz} state for their Suzaku 
and XMM spectra in the Suzaku band (0.3-0.7 keV). Besides the two oxygen lines we analyze here, which are the most important 
in the range 0.5-0.7 keV, the main ion lines dominating the Suzaku band, are due to CX collisions of $C^{6+}$, $N^{7+}$ 
and $N^{6+}$ with IS neutrals. These ions, although not treated in this study, are known to have equivalent abundance 
variations as $O^{7+}$ and $O^{8+}$ between solar maximum and solar minimum, and thus, an equivalent trend in the 
SWCX X-ray emissions they generate (Koutroumpa et al., 2006). 

The simulated O VII and O VIII intensities we derive are of the same order and even slightly higher than the ones attributed to the LB 
by \cite{henley07suz}. We conclude that, as in the case of the XMM data, the foreground emission in the 0.5 - 0.7 keV band 
can be exclusively attributed to the Heliosphere. We must also note, that according to $O^{7+}$ abundance 
measurements, the SWCX O VII emission is likely to have risen by a factor of 1.6 during the OFF exposure, with respect to the 
ON exposure. Therefore, the background emission attributed to the absorbed distant component should be revised, 
by subtracting properly the heliospheric contribution (see table \ref{targets}).

\subsection{Hubble Deep Field North}\label{HDFN}
The Hubble Deep Field-North has been observed with XMM-Newton as a test case for the noncosmic background modeling and subtraction for 
the European Photon Imaging Camera. The total length of the observation was broken in four exposures scheduled over a period of 16 days. 
The fourth pointing, on June 1, 2001 offered an exemplary case of SWCX emission detection, which was analysed by \cite{snowden04}.

The authors, in a useful exposure time of 38.1 ks out of a total 95.4 ks, after removal of flaring periods, discerned a drop in the 
0.52-0.75 keV band lightcurve, occuring in the last quarter of the interval. They used a separate spectral fit for the HIGH 
and LOW regime, as they define the intervals before and after the drop respectively, with four main components: (i) an unabsorbed 
thermal component with $T\sim 0.1$keV for the LB, (ii) an absorbed thermal component with also $T\sim 0.1$keV for the 
lower halo emission, (iii) an absorbed, hotter component ($T\sim 0.6$keV) for the Galactic halo or Local Group emission, 
and (iv) an absorbed power law with spectral index of 1.46 for the unresolved extragalactic sources. The absorption was fixed to the 
Galactic column density $N_H = 1.5\times 10^{20}$cm$^{-2}$. In the HIGH state of the exposure only, a series of spectral lines 
to represent the SWCX emission was added in the fitting process. The lines included were the C VI 0.37 and 0.46 keV lines, 
the O VII triplet at 0.57 keV, the O VIII lines at 0.65 and 0.81 keV, the Ne IX line at 0.91 keV and the Mg XI line at 1.34 keV. 
Their analysis yielded a SWCX induced line flux of 7.39 LU and 6.54 LU for the O VII and O VIII lines respectively in the HIGH state 
only, while the contribution of SWCX in the LOW state regime and the remaining pointing was completely neglected.

\setlength{\extrarowheight}{1mm}
\begin{table*}
\centering
\caption[\small{\textit{Parametres des simulations HDFN}}]
{\footnotesize{Simulation parameters and results for the HDFN observations. We list SW parameters, and relative abundances 
for the two variants of the simulation, as well as model and data line fluxes for O VII and O VIII.}\label{HDFN_table}}
\begin{minipage}[t]{\linewidth}
\begin{tabular}{ccccccccc}\hline \hline
& \multicolumn{4}{c}{\underline{\,\, Proton flux Step Function\,\, }} 
	& \multicolumn{4}{c}{\underline{\,\, Ion Relative Abundances\,\, }}\\
&\multicolumn{2}{c}{\underline{\,\, Simulation 1\,\, }} & \multicolumn{2}{c}{\underline{\,\, Simulation 2\,\, }} 
	& \multicolumn{2}{c}{\underline{\,\, HIGH\,\, }} & \multicolumn{2}{c}{\underline{\,\, LOW \,\, }} \\
&$T_o + \Delta t$ (d)& $f_{SW}$ (SPFU) & $T_o + \Delta t$  (d) & $f_{SW}$ (SPFU) &$\left[ O^{7+}\,/\,O\right] $ 
	& $\left[ O^{8+}\,/\,O\right] $ & $\left[ O^{7+}\,/\,O\right] $ & $\left[ O^{8+}\,/\,O\right] $ \\\hline 
& (0.5 + 0.5) & 4 & (0.5 + 1.) & 10 & 0.488 & 0.28 & 0.125 & 0.038 \\\hline \hline

&\multicolumn{8}{c}{\underline{\,\, Line flux (LU = photons cm$^{-2}$ s$^{-1}$ sr$^{-1}$)\,\, }}\\
 & \multicolumn{2}{c}{\underline{\,\, Simulation 1\,\, }} & \multicolumn{2}{c}{\underline{\,\, Simulation 2\,\, }} 
	& \multicolumn{2}{c}{\underline{\,\, SCK04\,\, }} & \multicolumn{2}{c}{\underline{\,\, Data\,\, }}\\
 & High & Low & High & Low & \multicolumn{2}{c}{High} & High & Low \\\hline 
O VII & 4.20  & 3.22 & 9.06 & 2.02 & \multicolumn{2}{c}{7.39} & 15.44 & 8.14 \\
O VIII & 1.99 & 1.40 & 5.69 & 1.18 & \multicolumn{2}{c}{6.54} & 8.14 & 1.96 \\\hline \hline
\end{tabular}
\end{minipage}
\end{table*}

In section 5 of \cite{koutroumpa06}, we presented a first time-dependent modeling of the SWCX emission, based on the same method 
presented in this study and applied on the HDFN June 1, 2001 exposure. We have used a simplified geometry of the LOS, supposing 
that it was lying in the ecliptic plane instead of pointing at a helioecliptic latitude of 57\deg . We have presented simulated 
lightcurves for two different sets of SW parameters and concluded that the heliospheric X-rays were responsible for the observed 
enhancement and mainly that the LOW state spectrum still contained a large fraction of heliospheric emission. Unfortunately, 
an error in taking into account the variable steps on the LOS, which was discovered after publication annuls our results, but not
our main conclusions which we attempt to validate in the present re-analysis of the HDFN simulation.
 
We performed two simulations, taking into account the real geometry of the HDFN observation (Observer at 
$\lambda_{obs}$ = 251\deg , LOS pointing backwards at $\lambda$, $\beta$ = 48$^\circ$, 57$^\circ$, ) 
and considering $O^{7+}$ and $O^{8+}$ relative abundances as implied by the O VIII and O VII lines ratio calculation 
in \cite{snowden04} (\cite{snowden04}; \cite{koutroumpa06}, see table \ref{HDFN_table}). The difference in the two simulations 
lies on the step function adopted for the SW enhancement and how it affects the high latitude LOS of the HDFN. 

\begin{figure*}
\centering
\subfigure[Simulation 1]
{
\label{hdfn_real}
\includegraphics[width=0.48\textwidth ]{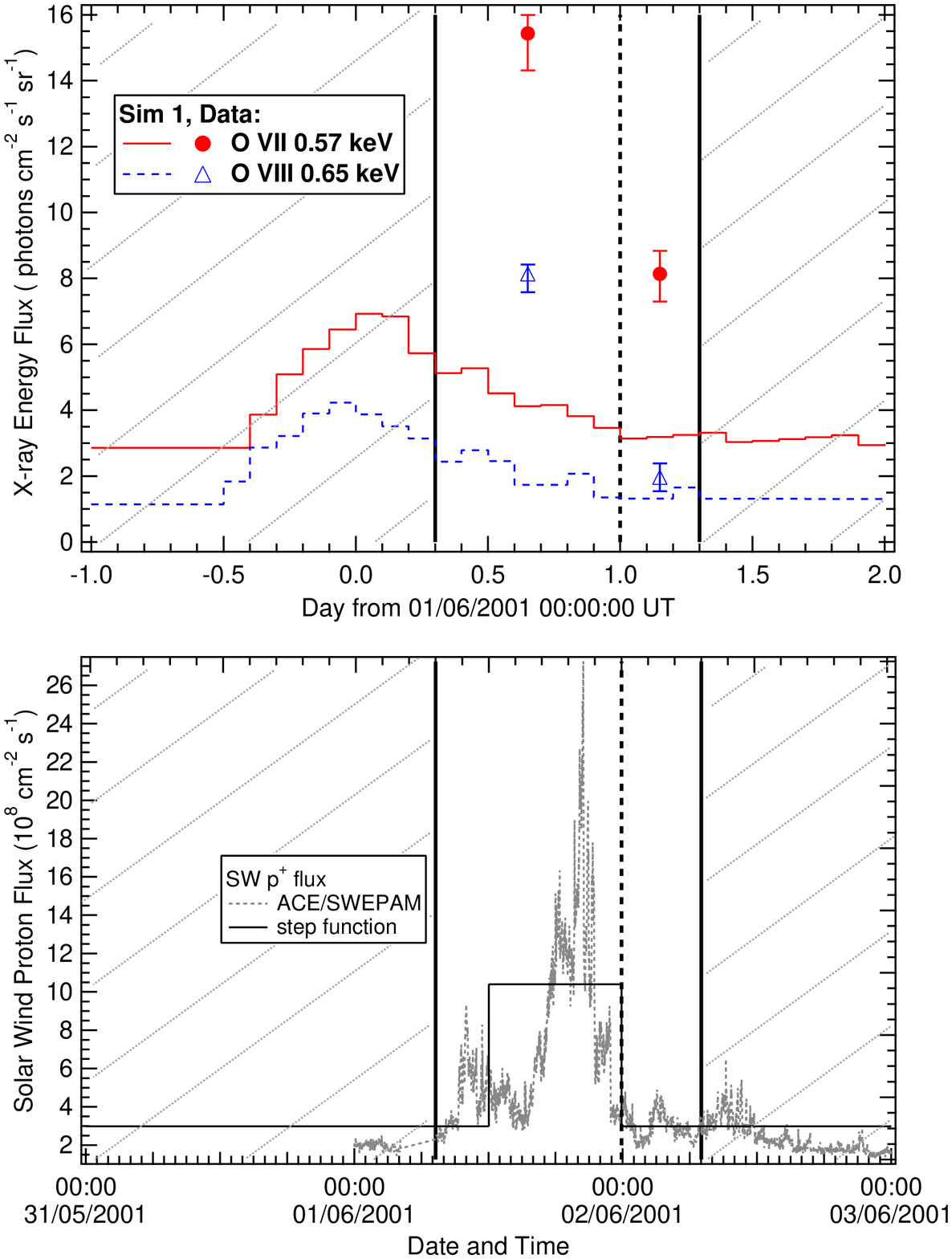}
}
\subfigure[Simulation 2]
{
\label{hdfn_perf}
\includegraphics[width=0.48\textwidth ]{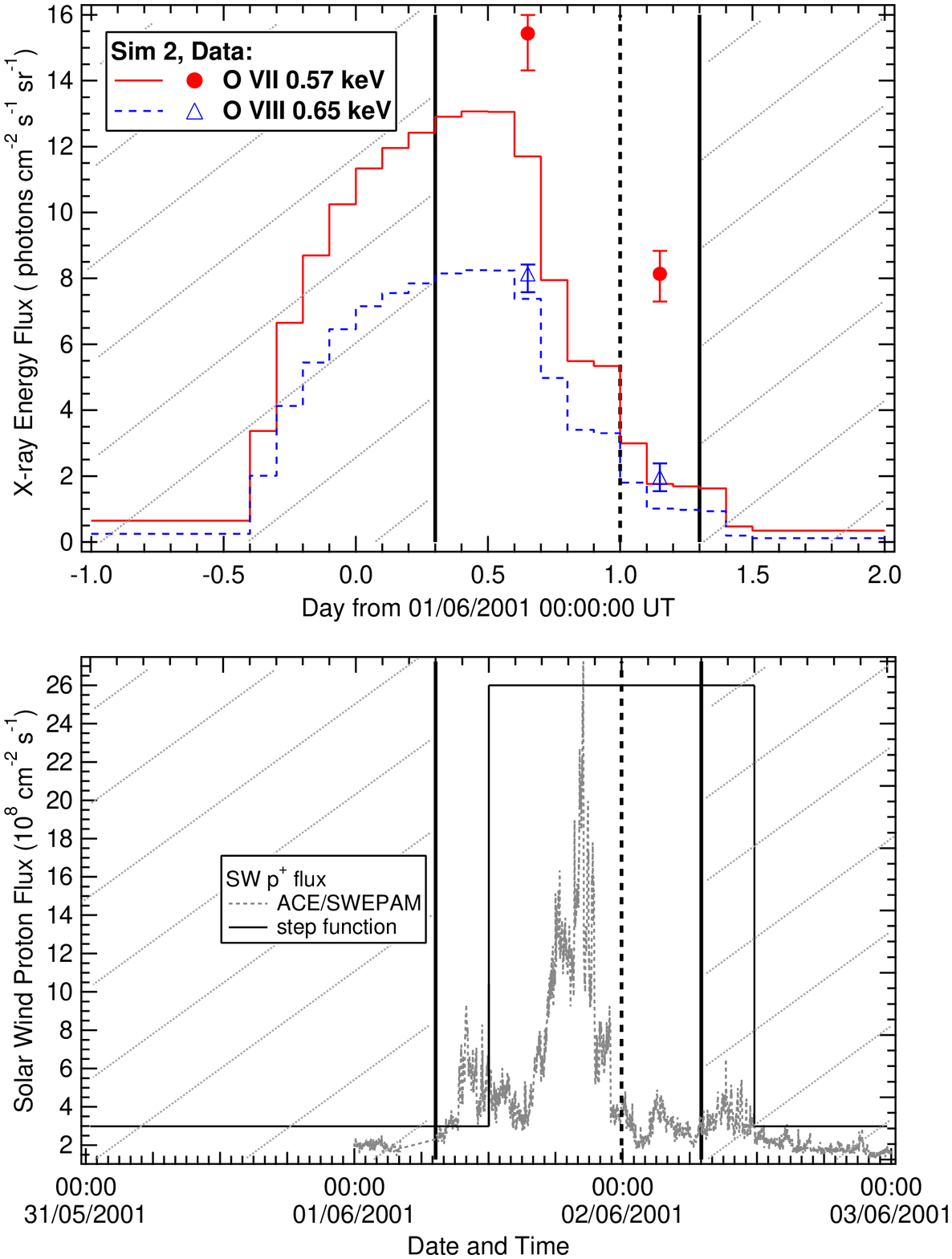}
}
\caption{\footnotesize{Solar Wind input conditions, SWCX simulation results and XMM-Newton data analysis results 
for the HDFN June, 1, 2001 observation. \textit{Column (a)}: Simulation 1 based on real-time SW measurements in the ecliptic, 
\textit{Column (b)}: Simulation 2 based on best model-XMM data correlation. 
\textit{Upper Panels}: Simulated lightcurves for the O VII and O VIII line emission 
in Line Units. The plain red line is for the O VII emission and the dashed blue line is for the O VIII emission. 
Red circles represent the XMM-measured O VII line fluxes, while the blue triangles represent 
the XMM-measured fluxes for O VIII line. 
\textit{Lower Panel}: Solar wind proton flux (dotted line) in units of $10^8$ cm$^{-2}$ s$^{-1}$ for the same period. 
The step function simulating the SW enhancement in each simulation is presented with the plain black line.
In all panels, the vertical plain lines represent the start and end of the observation period, 
while the dashed vertical line is the separation between the High and Low regimes as defined in \cite{snowden04}.}}
\label{hdfn_lcurve}
\end{figure*}

In the first case, we take equivalent conditions as those presented in \cite{koutroumpa06}, with a step function occuring on day 
$T_o + \Delta t$ = (0.5 + 0.5) d, and a SW proton flux of 4 SPFU. The results are presented in the upper panel of figure 
\ref{hdfn_real}. 

We can see from figure \ref{hdfn_real} that timing between model and data seems correct, as it did for the simplified geometry used 
in \cite{koutroumpa06}. The simulated lightcurve, within the observation limits, starts at an already higher value, dropping at the 
end of the SW enhancement. Nevertheless, the absolute difference between HIGH and LOW regimes in the simulation is far too low 
to explain the large drop measured in XMM data. Indeed the HIGH to LOW drop is of 0.98 LU and 0.59 LU only, in the model for the line O VII 
and O VIII respectively, while it is of 7.3 LU for O VII and of 6.18 for O VIII in the data.

The HDFN is the only exposure divided in two states. Since the target field is the same for the two states, the column density is 
identical, thus, we are confident that such a variation in the soft X-ray background is mainly due to the heliospheric emission. 
The Geocorona could partly be responsible for the variation, but as explained in \cite{cravens01} and \cite{snowden04} 
it should be temporally correlated with the SW rising and dropping with it. Moreover, SCK04 analyze the 
XMM-Newton geometry for the 01/06/2001 and conclude that the satellite avoids the main zones that could be contaminated by 
the Geocorona. 

As a consequence, we wished to test our SWCX model and see if we can find reasonable SW enhancement conditions that could fit better 
this strong variation in the Soft X-ray background. We consider then, a step function occuring on $T_o + \Delta t$ = (0.5 + 1.0) d and 
with a SW proton flux of 10$\times$\SPFU , which is quite rare, but not impossible to occur. We assume that the AR producing 
this enhancement is not infinitely ancient on the solar disk, as assumed in the general time-dependant modeling described 
in Sect.\ref{SWCXmodel}, but appears $\sim$7d earlier than 01/06/2001 00:00UT (day 0), while pointing radially at a helioecliptic 
longitude of $\lambda_j \sim$219\deg . 

Under such conditions, the LOS will be affected only up to the segment 
($\lambda_j , \beta_j \sim$ 219\deg , 40\deg ) which lies at phase angle $\phi_j \sim$ 32\deg\ and distance 1.08 AU 
with respect to the observer (see fig.\ref{losback} and Sect.\ref{SWCXmodel}). We, therefore, assume that the SW leaving the Sun 
on day -7d and affecting the major emitting part of the LOS was quite different than the one measured on day 0.5 
at the Earth's position. This is a plausible hypothesis, since the SW variation time-scales can be much shorter than 7 days. 
In addition, we cannot be certain for the SW conditions at a $\sim$30\deg\ angular distance from the Earth, since there are no 
in situ measurements at such distances.

The results of this simulation are presented in figure \ref{hdfn_perf} with the same annotations as in figure \ref{hdfn_real}. 
The 'truncation' of the spiral's effect on the LOS at 40$^{\circ}$ forces a harder drop between the HIGH and LOW states, 
which correspond better to the observed temporal profile. Moreover, the increase of the SW proton flux we imposed allows a 
better comparison between the simulated and measured HIGH-LOW differences. Indeed, the drop of the O VII line flux is of 7.04 LU. 
The O VIII line drops by 4.51 LU which is still a little low with respect to the data drop, but we could suppose that ion relative 
abundances are also very different than the ones measured in Earth orbiting solar instruments.

What we need to stress as a conclusion to these simulation tests we performed, is the importance of accurate and multi-positioned 
in-situ measurements of the SW properties. Since SWCX emission is so sensitive to SW flux and abundance variations, precise simulations 
can only be improved with detailed and liable input parameters.

\subsection{Marano Field}\label{marano}
The Marano Field was named by an optical quasar survey by \cite{marano88} and was frequently surveyed in optical wavelengths, 
X-rays (ROSAT, XMM) and radio wavelengths (\cite{krumpe07} and references within). The mean galactic H I column density in the Marano 
Field is $N_H$ = 2.7$\times$10$^{20}$ cm$^{-2}$. XMM-Newton observed the Marano Field 7 times 
in the period between August 22, and August 27, 2000. The mean total exposure time was $\sim$10 ks for each observation.

The SWCX model ground level is taken for maximum solar activity for 2000. Marano field is pointing at -67\deg of helioecliptic 
latitude which means the SW affecting the outer parts of the LOS could have been very different than the one measured in the ecliptic.

The SW variations measured form WIND instrument at this time period can be modeled as 3 step functions starting on days 
(from date 22/08/2000 00:00UT): 0 d (step-1), 1.625 d (step-2) and 5 d (step-3) lasting 1.4 d, 0.4 d and 1.8 d respectively. 
Step-1 yields an enhancement of 2.12$\times$\SPFU, step-2 an enhancement of 2.7$\times$\SPFU and finaly step-3 gives an enhancement 
of 1.8$\times$\SPFU . There are ACE/SWICS measurements showing high values of $O^{8+} = 0.14$ and $O^{7+} = 0.3$ abundances before 
and during step-1 and step-2.

In figure OM-\ref{marano_lcurve}, we present the lightcurves for O VII and O VIII lines resulting from the simulation, 
as well as the XMM-data for the 7 Marano exposures. The variations measured with XMM in the Marano field are not very large, 
mainly inside the error bars of each data point. The model predicts relative variations of the same order and with the 
right timing as well. We see a moderate mean increase around day 2 (24/08/2000) of about 38\% in the O VII line, 
which is reproduced with a relative strength of 34\% in the model. The O VIII data points have larger 
uncertainties, but the mean relative increase in the exposures period is around 50 - 80\% (considering the points with 
the smallest error bars), while the model predicts a relative increase of 40\% .

\subsection{Lockman Hole}\label{lockman}
The Lockman Hole is the sky region with the absolute lower H I column density ($N_H \approx 4.5\times 10^{19}$ cm$^{-2}$). 
XMM-Newton observed the Lockman Hole 7 times in the period between October 15, and October 28, 2002. The total exposure time 
spans from $\sim$80 ks to $\sim$105 ks for each observation.

The SWCX model ground level is taken for maximum solar activity, 2002 being closer to maximum. 
To model the Lockman Hole exposures during the period 15 - 28/10/2002 we use two step functions starting at days -9.25 (step-1) 
and 0 (step-2). Step-1 has a SW proton flux of 1.7$\times$\SPFU\ lasting for about 2 days, while step-2 has a proton flux 
of $\sim$ 2$\times$\SPFU\ lasting for about 2 days. $O^{7+}$ relative abundances could be slightly enhanced during the two 
SW enhancements ($\sim$ 0.3 - 0.4, although not used in the figure), but $O^{8+}$ abundances are very difficult to estimate 
because of big lack of data. The Lockman Hole is pointing at an helioecliptic latitude of 45\deg\ which could allow very different 
SW caracteristics affecting the LOS than those measured in Solar instruments.

The resulting lightcurves are presented in figure OM-\ref{lock_lcurve}. Clearly besides the discrepancies in 
the variation amplitudes, we also have a timing problem in the comparison, 
the model predicting an emission rising too fast with respect to the measured enhancement for step-1 and too late 
for step-2 respectively. These discrepancies do not have any obvious explanation but an inherent variation in the O VII and O VIII 
emission of the Lockman Hole field (more discussion in \S \ref{xmmVSswcx}).

\setlength{\extrarowheight}{2mm}
\begin{table*}
\begin{center}
\caption[\small{\textit{Comparaison spectrale des modèles aux observations XMM}}]
{\footnotesize{List of XMM-Newton re-processed observations. Comparison of total SXRB O VII and O VIII line intensity 
in XMM/PLC data fit (see \S \ref{xspec}) and SWCX model results.}\label{targets}}
\begin{minipage}[t]{\linewidth}
\begin{tabular}{lcl@{}ccccccc}\hline \hline

\multicolumn{4}{c}{} &\multicolumn{6}{c}{\underline{\,\, Line flux (LU = photons cm$^{-2}$ s$^{-1}$ sr$^{-1}$)\,\, }}\\ 
\multicolumn{2}{c}{\underline{\,\, Target\,\, }} & \multicolumn{2}{c}{} & \multicolumn{2}{c}{\underline{\,\, Data (Total)\,\, }} 
	& \multicolumn{2}{c}{\underline{\,\, CX Model\,\, }}& \multicolumn{2}{c}{\underline{\,\, Residual\,\, }}\\ 
Name & Gal.Coord. & Obs ID & Start Date
	& O \tiny{VII} & O \tiny{VIII} & O \tiny{VII} & O \tiny{VIII}& O \tiny{VII} & O \tiny{VIII}\\\hline

\multirow{7}{*}{\begin{sideways}Marano Field\end{sideways}} & \multirow{7}{20mm}{(269.8$^\circ$,-51.7$^\circ$)} 
	& x:129320801 & 22/08/2000 & 6.67$^{+1.26}_{-1.01}$ & 2.22$^{+0.66}_{-0.53}$ & 1.46 & 0.58 & 5.21 & 1.64\\
&  & x:129320901 & 22/08/2000 & 5.55$^{+1.08}_{-0.92}$ & 1.97$^{+0.50}_{-0.45}$ & 1.49 & 0.59 & 4.06 & 1.38\\
&  & x:110970201 & 24/08/2000 & 5.58$^{+1.52}_{-1.02}$ & 2.91$^{+0.65}_{-0.41}$ & 1.80 & 0.74 & 3.78 & 2.17\\
&  & x:110970301 & 24/08/2000 & 6.82$^{+1.29}_{-1.32}$ & 2.43$^{+0.60}_{-0.60}$ & 1.87 & 0.78 & 4.95 & 1.65\\
&  & x:110970401 & 24/08/2000 & 7.68$^{+1.61}_{-1.35}$ & 1.60$^{+0.76}_{-0.62}$ & 1.91 & 0.80 & 5.77 & 0.80\\
&  & x:110970501 & 26/08/2000 & 6.17$^{+1.14}_{-0.88}$ & 2.03$^{+0.50}_{-0.42}$ & 1.50 & 0.56 & 4.67 & 1.47\\
&  & x:110970701 & 30/08/2000 & 7.25$^{+1.16}_{-0.78}$ & 1.16$^{+0.55}_{-0.37}$ & 1.36 & 0.43 & 5.89 & 0.73\\\hline 

\multirow{2}{*}{\begin{sideways}\,HDFN\,\end{sideways}}& \multirow{2}{20mm}{(126.0$^\circ$,55.2$^\circ$)} 
& \multirow{2}{18mm}{x:111550401} & 01/06/2001 \scriptsize{HIGH} & 15.44$^{+0.56}_{-1.12}$ & 8.14$^{+0.28}_{-0.56}$ &  9.06 & 5.69 
	& 6.38 & 2.45\\
& & & 01/06/2001 \scriptsize{LOW} & 8.14$^{+0.70}_{-0.84}$ & 1.96$^{+0.42}_{-0.42}$ & 2.02 & 1.18 & 6.12 & 0.78\\\hline	 

\multirow{7}{*}{\begin{sideways}Lockman Hole\end{sideways}} & \multirow{7}{20mm}{(149.1$^\circ$,53.6$^\circ$)} 
	& x:147510101 & 15/10/2002 & 8.76$^{+1.80}_{-0.74}$ & 2.30$^{+0.68}_{-0.53}$ & 2.46 & 0.96 & 6.30 & 1.34\\
& & x:147510801 & 17/10/2002 & 18.12$^{+1.84}_{-1.62}$ & 3.40$^{+0.79}_{-0.73}$ & 2.30 & 0.90 & 15.82 & 2.5\\
& & x:147510901 & 19/10/2002 & 16.69$^{+3.17}_{-1.72}$ & 3.90$^{+1.73}_{-0.83}$ & 2.10 & 0.82 & 14.59 & 3.08\\
& & x:147511001 & 21/10/2002 & 10.69$^{+1.42}_{-1.14}$ & 1.80$^{+0.66}_{-0.47}$ & 1.99 & 0.78 & 8.70 & 1.02\\
& & x:147511101 & 23/10/2002 & 18.47$^{+3.95}_{-1.60}$ & 3.91$^{+1.99}_{-0.72}$ & 2.11 & 0.83 & 16.36 & 3.08\\
& & x:147511201 & 25/10/2002 & 7.25$^{+1.81}_{-1.45}$ & 2.05$^{+0.89}_{-0.72}$ & 2.47 & 0.96 & 4.78 & 1.09\\
& & x:147511301 & 27/10/2002 & 13.94$^{+2.54}_{-1.60}$ & 2.61$^{+1.01}_{-0.97}$ & 2.44 & 0.95 & 11.50 & 1.66\\\hline 
	
\multirow{2}{*}{\begin{sideways}Filament\end{sideways}}& (278.7$^\circ$,-45.3$^\circ$) 
	& x:084960201 & 03/05/2002 \scriptsize{ON} & 11.38$^{+1.51}_{-1.65}$ & 3.36$^{+0.73}_{-0.70}$ & 3.16 & 1.02 & 8.96 & 2.37 \\
& (278.7$^\circ$,-47.1$^\circ$) 
	& x:084960101 & 03/05/2002 \scriptsize{OFF}&  16.95$^{+2.66}_{-2.67}$ & 2.74$^{+1.10}_{-1.06}$ & 3.47 & 1.11 & 11.72 & 1.18\\\hline\hline 
\end{tabular}
\end{minipage}
\end{center}
\end{table*}

\section{Discussion}\label{discussion}
\subsection{Non-shadowed regions}\label{xmmVSswcx}
In table \ref{targets} we resume all the XMM observation fields modeled in the SWCX simulations. We include the 
SG Filament XMM ON exposure, since it is a minor absorber and substantial emission of the halo remains. The data results 
are for the total O VII and O VIII line intensity as calculated in the XMM/PLC spectral model we defined in \S \ref{xspec}. 
The data contain, thus, both local foreground and 
distant components of the O VII, O VIII diffuse emission. Model results stand for the heliospheric O VII and O VIII line emission, 
ONLY, due to charge-exchange collisions. Finally in the last two columns we present the residual diffuse SXRB emission when 
we subtract the SWCX component from the total data results. This residual emission, should be originating in the Galactic Halo 
(or the border regions of the Loop I), 
since from what we already demonstrated in the shadowing results, the LB emission can be accounted as exclusively 
CX-induced heliospheric emission (see also following \S \ref{shadowsVSswcx}).

The residual emission is quite variable, since the SWCX temporal model did not succeed in correlating temporally all series 
of observations. Notably, the Lockman Hole variations are due to the complete lack of temporal correlation between 
the observed X-ray emission variations and the SWCX model predictions. These variations, we do not manage to explain, unless 
there is an inherent variation in the Galactic halo emission between the different Lockman Hole exposures. But, 
according to the absorption law: 
\begin{equation}
I_{OVII} \propto exp(-\sigma_{OVII}\, N_H)
\end{equation}
where $\sigma_{OVII}\sim8.56\times10^{-22}$ cm$^2$ is the photoelectric absorption cross section for O VII energies 
(\cite{morrison83}), to have a factor of 2 decrease in O VII line intensity, we need a factor of $\sim$20 increase in 
the H I column density $N_H$, which is contested in the H I galactic atlas of \cite{hartmann97}. In addition, \cite{kappes03} 
demonstrated that in the Lockman Hole region of the sky the halo is quite homogeneous. It is therefore probable 
that the Lockman observations were contaminated by solar protons or terrestrial CX emission.

For the HDFN field we have retained the simulation 2 results of the SWCX model, since they are the ones that give 
the most stable residual emission as required for a unique long-lasting exposure towards a constant field (see \S \ref{HDFN}). 
The Marano field is the only one to have relatively constant residual emission, since already data were quite 
stable even before the CX component subtraction.
 
We understand that, in general, XMM data comparison to model results provides quite poor correlation, especially for the O VII line 
intensity. The data are systematically higher than model predictions, by factors ranging from 2 to 10. This is expected, naturally, 
since the model accounts only for the CX heliospheric component of the SXRB. In addition, the residual emission, after the 
CX subtraction is quite different for each target region. This is due to the different H I column density in each field, or, to 
intrinsic variations of the Galactic Halo emission from one region to another. 

The O VII data points (and thus the residual intensity) exhibit a higher dispersion than the O VIII data points for each target. 
Except for the HDFN which, as we explained in \S\ref{HDFN}, was temporally divided in two exposures towards the exact same field, 
all the other targets, were observed in repeated exposures in which the central field of view was pointing in slightly different 
directions. In the O VIII line energies, the transmission is very high, larger than 60\%\ for column densities less than 
$\sim$10$^{21}$ (\cite{freyberg04b}; \cite{henley07xmm}), like our XMM targets, and thus is less sensitive to column density 
variations. However, the O VII line is much more absorbed, so even minor variations of the $N_H$ column density induce large 
dispersion in the emission measured for the repeted exposures at each galactic region.
 
In the future, we need to pursue the study with better statistics for each target field, i.e. apply the SWCX model to more 
exposures towards each field so as to fit separately each target results.

\subsection{Shadowing observation data vs SWCX model}\label{shadowsVSswcx}
The most important result in our analysis is the one deriving from the shadowing observations, which is that the emission 
initially attributed to the Local Interstellar Bubble till now, is probably originating entirely from within the Heliosphere. 

\begin{figure}
\centering
\includegraphics[width=0.5\textwidth ,height=!]{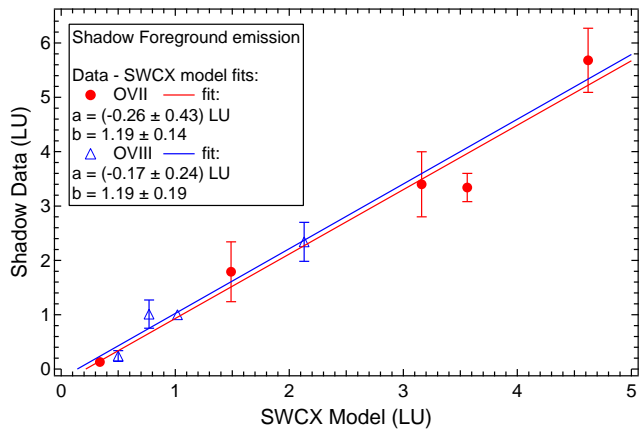}
\caption{\footnotesize{Foreground emission data from shadowing observations in litterature (MBM 12: \cite{smith05}; 
\cite{smith07} and SGF: \cite{henley07xmm}; \cite{henley07suz}) versus SWCX model results 
for the O VII and O VIII line intensities. Linear fits are calculated with no error-bar weighting, since not 
all data error-bars were communicated (see table \ref{shadows} and \S \ref{shadowsVSswcx} for details).}}
\label{shadow_fit}
\end{figure}

In table \ref{shadows}, constructed in the same principle as table \ref{targets}, we resume the MBM 12 and SG Filament shadowing 
observations. In figure \ref{shadow_fit} we linearly fit the data over the SWCX model results for each shadowing observation. 
We did not use data errorbars as 
standard deviation to weight the fit, because not all data errorbars were communicated (\cite{henley07xmm} and Henley et al., 2007b). 
The fit coefficients (y(LU) = (a + bx)(LU)) for the O VII and O VIII lines are also noted on the figure. 

The observations data represent the LB O VII and O VIII line intensities as derived from the authors (\cite{smith05}; Smith et al., 2007; 
\cite{henley07xmm}; and Henley et al., 2007b). As explained individually for each case in sections \ref{mbm12} and \ref{filament}, 
the shadows block more or less efficiently the Galactic Halo oxygen emission, and in their analysis the authors derive only 
the local foreground emission of the oxygen lines. 

The only exception is in the Suzaku/MBM 12 observation (Smith et al., 2007) where the O VII and O VIII halo emission is added 
to the foreground emission for the OFF-CLOUD exposure. However, as we demonstrated in \S\ref{mbm12_suz}, there is 
a $\sim$30\% and $\sim$55\% increase in the OFF-CLOUD simulated SWCX O VII and O VIII line intensities respectively, 
due to the brief SW enhancement at the end of the ON-CLOUD exposure. \cite{smith07} supposed that the foreground (LB) emission 
was constant between the ON and OFF cloud exposures, thus this increase was erroneously attributed to the Galactic Halo emission 
which should be revised. For this reason, we include these values in the linear data-model fits as well.  

The linear fit has a slope of 1.19 ($\pm$ 0.14; 0.19 (1$\sigma$) for O VII and O VIII resp.) which means that the SWCX model 
reproduces very well the local foreground emission measured in shadows. Moreover, the intersection of the linear fit with the 
data axis should give an estimate of the residual foreground emission to be attributed to the LB. The foreground is found to be 
(-0.26 $\pm$ 0.43 (1$\sigma$)) LU for O VII and (-0.17 $\pm$ 0.24 (1$\sigma$)) LU for O VIII, which means that with a high probability 
the LB O VII and O VIII emission is negligible compared to the heliospheric emission. 

\setlength{\extrarowheight}{1mm}
\begin{table*}
\begin{center}
\caption[\small{\textit{Comparaison spectrale des modèles aux observations XMM}}]
{\footnotesize{List of shadowing observations. Comparison of previous data-analyses results with our SWCX model. 
Data results include mainly the LB O VII and O VIII line fluxes. $^a$Only \cite{smith07} OFF exposure with Suzaku 
includes O VII and O VIII strengths of LB and halo emission (see \S \ref{mbm12_suz} for details). 
$^b$ Prefixes are x: for XMM-Newton, c: for CHANDRA, s: for SUZAKU. $^{c1}$\cite{smith05}, $^{c2}$\cite{smith07}. 
$^{d1}$\cite{henley07xmm} \& $^{d2}$\cite{henley07suz} see $\S$\ref{filament} for details.}\label{shadows}}
\begin{minipage}[t]{\linewidth}
\begin{tabular}{lclccccccc}\hline \hline
\multicolumn{4}{c}{} &\multicolumn{6}{c}{\underline{\,\, Line flux (LU = photons cm$^{-2}$ s$^{-1}$ sr$^{-1}$)\,\, }}\\ 
\multicolumn{2}{c}{\underline{\,\, Target\,\, }} & \multicolumn{2}{c}{} & \multicolumn{2}{c}{\underline{\,\, Data (LB$^a$)\,\, }} 
	& \multicolumn{2}{c}{\underline{\,\, CX Model\,\, }}& \multicolumn{2}{c}{\underline{\,\, Residual\,\, }}\\ 
Name & Gal.Coord. & Obs ID$^b$ & Start Date
	& O \tiny{VII} & O \tiny{VIII} & O \tiny{VII} & O \tiny{VIII}& O \tiny{VII} & O \tiny{VIII}\\\hline

\multirow{3}{*}{\begin{sideways}\,MBM12\,\end{sideways}} & (159.2$^\circ$,-34.5$^\circ$) 
	& c:900015943 & 17/08/2000  \scriptsize{ON}& 1.79$\pm$0.55$^{c1}$ & 2.34$\pm$0.36$^{c1}$	& 1.49 & 2.13 &  0.30 & 0.21\\
& (159.2$^\circ$,-34.5$^\circ$) 
	& s:500015010 & 03/02/2006 \scriptsize{ON} & 3.34$\pm$0.26$^{c2}$ & 0.24$\pm$0.10$^{c2}$ & 3.56 & 0.50 & 0:: & 0:: \\
& (157.3$^\circ$,-36.8$^\circ$) 
	& s:501104010 & 06/02/2006 \scriptsize{OFF} & 5.68$\pm$0.59$^{c2}$ & 1.01$\pm$0.26$^{c2}$ & 4.62 & 0.77 & 1.06 & 0.24\\\hline 

\multirow{3}{*}{\begin{sideways}Filament\end{sideways}}& (278.7$^\circ$,-45.3$^\circ$) 
	& x:084960201 & 03/05/2002 \scriptsize{ON} & 3.4$^{+0.6}_{-0.4}$ $^{d1}$& 1 $^{d1}$	& 3.16 & 1.02 & 0.24 & 0:: \\
& (278.7$^\circ$,-47.1$^\circ$) 
	& x:084960101 & 03/05/2002 \scriptsize{OFF}&   &  & 3.47  & 1.11  & 0:: & 0::\\
& (278.7$^\circ$,-47.1$^\circ$) 
	& s:501001010 & 01/03/2006 \scriptsize{ON}& 0.13$^{d2}$ & N.A.$^{d2}$  & 0.34 & 0.02:: & 0:: & 0::\\\hline\hline 
\end{tabular}
\end{minipage}
\end{center}
\end{table*}

\subsection{The ROSAT 3/4 keV map}\label{rosat}
\begin{figure}
\centering
\includegraphics[width=0.5\textwidth ,height=!]{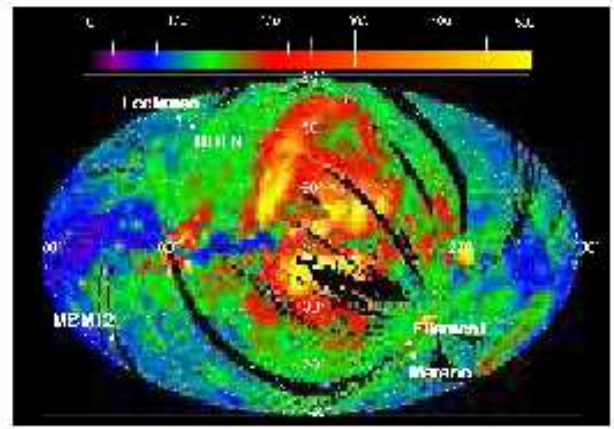}
\caption{\footnotesize{ROSAT 3/4 keV all-sky map (\cite{wang98}) in ROSAT units (10$^{-6}$ cts s$^{-1}$ arcmin$^{-2}$) 
including target fields modeled with the SWCX model.}}
\label{ROSAT075}
\end{figure}

In figure \ref{ROSAT075} we present a 3/4 keV all-sky map of the ROSAT survey in galactic coordinates, extracted from \cite{wang98}. 
The map colorscales are in ROSAT units (1 RU = 0.07 LU). On the map we have marked the positions of the target fields 
we have presented in our analysis. The survey maps were constructed during a 6-month period, due to the constraints of the ROSAT 
observation geometry scanning the sky through a great circle perpendicular to the Sun-ROSAT direction. Data decontamination 
removed contamination due to LTEs to some extent, but not all of the heliospheric emission. 

North Galactic hemisphere fields have lower column density than the ones in the South Galactic hemisphere. HDFN, Lockman give 
an equivalent mean residual emission for the Halo, except for unexplicable variations of the Lockman Hole. The Marano Field 
should have similar residual emission as the SGF, since they lie in the same region, but the column density is different for Marano 
(\cite{krumpe07}) than the two SGF ON and OFF fields (\cite{henley07xmm}).

MBM 12 has the largest column density, especially compared to the other shadowing case we analyze, the SGF. Moreover, it is the 
only field pointing at no particular emissive feature in the 3/4 keV band. MBM 12 has only 23$\leq$ I $\leq$ 75 ROSAT units 
($10^{-6}$ counts s$^{-1}$ arcmin$^{-2}$), ON and OFF limits as measured by \cite{SMV93}. The equivalence between Line Units 
and ROSAT Units being 1RU = 0.07 LU, MBM 12 counts $\sim$1.6 LU ON-CLOUD and $\sim$5.25 LU OFF-CLOUD. The ROSAT observation 
of MBM 12 was also during a solar maximum, which is why the ON-CLOUD measurements are of the same order as the ON-CLOUD 
O VII measurements (1.79 LU) from the Chandra/MBM 12 observation (the OFF-CLOUD Chandra O VII; O VIII data were not communicated) 
and as the solar maximum SWCX model towards the MBM 12 (1.33 (ground level) - 1.49 LU). Summing the O VII and O VIII line 
intensities in the Chandra observations would highly exceed the ROSAT observations, but O VIII was highly contaminated 
by transient SWCX emission, as discussed in Sect.\ref{mbm12_chan}.

All the other fields are in the vicinity of the Loop I limits were and could be contaminated by the Loop's emission as well. 
The HDFN, Lockman Hole, SGF and Marano fields yield around 100 - 150 ROSAT units, i.e. 7 - 10.5 LU, which is of the 
same order of magnitude as the XMM O VII and O VIII data reported in table \ref{targets} (except HDFN which is highly contaminated 
by SWCX emission). The SWCX model predicts a heliospheric emission at least one fifth of the ROSAT emission maps in the brightest 
regions in the map, besides the Loop I, which leaves substantial contamination of the Heliosphere to the Galactic Halo emission. 

\section{Conclusions}\label{conclusion}	
For a series of deep field and shadow X-ray observations we have calculated the X-ray emission 
due to charge-exchange collisions in the Heliosphere, taking into account the temporal evolution of the emission due 
to solar cycle phase and/or solar wind enhancements. 

This analysis yields estimates of the SWCX heliospheric component within the diffuse Soft X-ray Background and 
confirms the large contamination of X-ray data by the heliospheric emission. 

The local 3/4 keV emission (due essentially to O VII and O VIII) detected in front of shadowing clouds is found to be 
\textbf{entirely} explained by the CX heliospheric emission. No emission from the LB is needed at these energies. 
Such a negative result will have to be taken into account in future determinations of the LB temperature and pressure. 

Galactic (halo + Loop I) oxygen line intensities need to be corrected for heliospheric contamination. We provide 
some corrections appropriate to the directions we have considered. Again, halo gas determinations must include 
these new constraints.

Our analysis is limited by the absence of out-of-ecliptic solar wind measurements and ionic abundances 
with enough statistics. Nevertheless, if pursued for a large sample of observations, it should better constrain 
the galactic emission. Future X-ray instruments with higher spectral resolution 
should allow a more precise determination of oxygen (and other) emission lines and better disentangle the different sources 
of emission.

\acknowledgements
We wish to thank John C. Raymond and Richard J. Edgar for useful information and enriching discussions. We would also like 
to thank our referee, Thomas E. Cravens, for his attentive report and valuable comments.

\Online

\appendix
\section{ \\Propagation of SW variations and sub-segment calculations in the SWCX model}\label{annexe}
\begin{figure*}
\centering
\subfigure[Forward LOS]
{
\label{losfront}
\includegraphics[width=0.45\textwidth ]{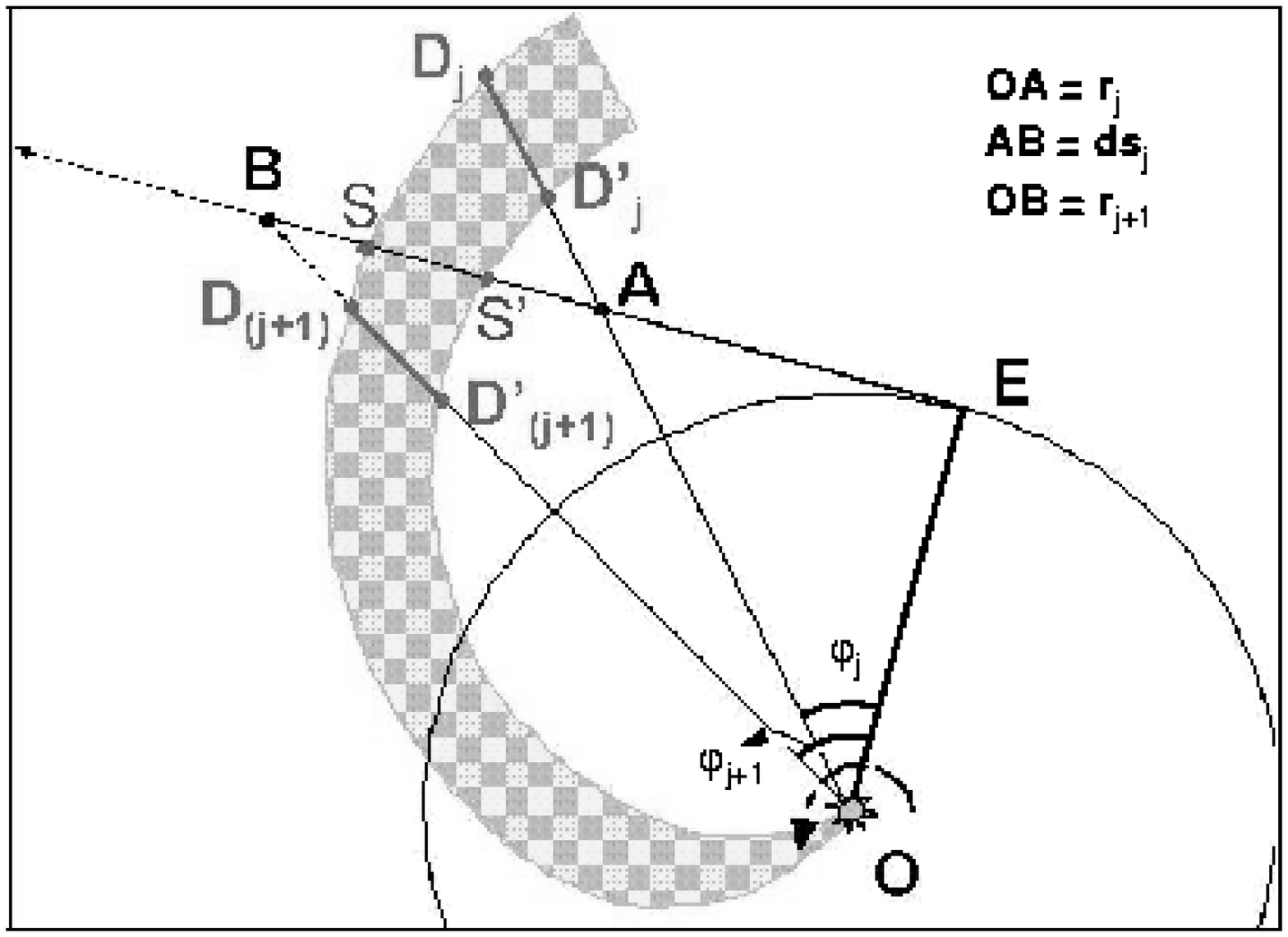}
}
\subfigure[Backward LOS]
{
\label{losback}
\includegraphics[width=0.45\textwidth]{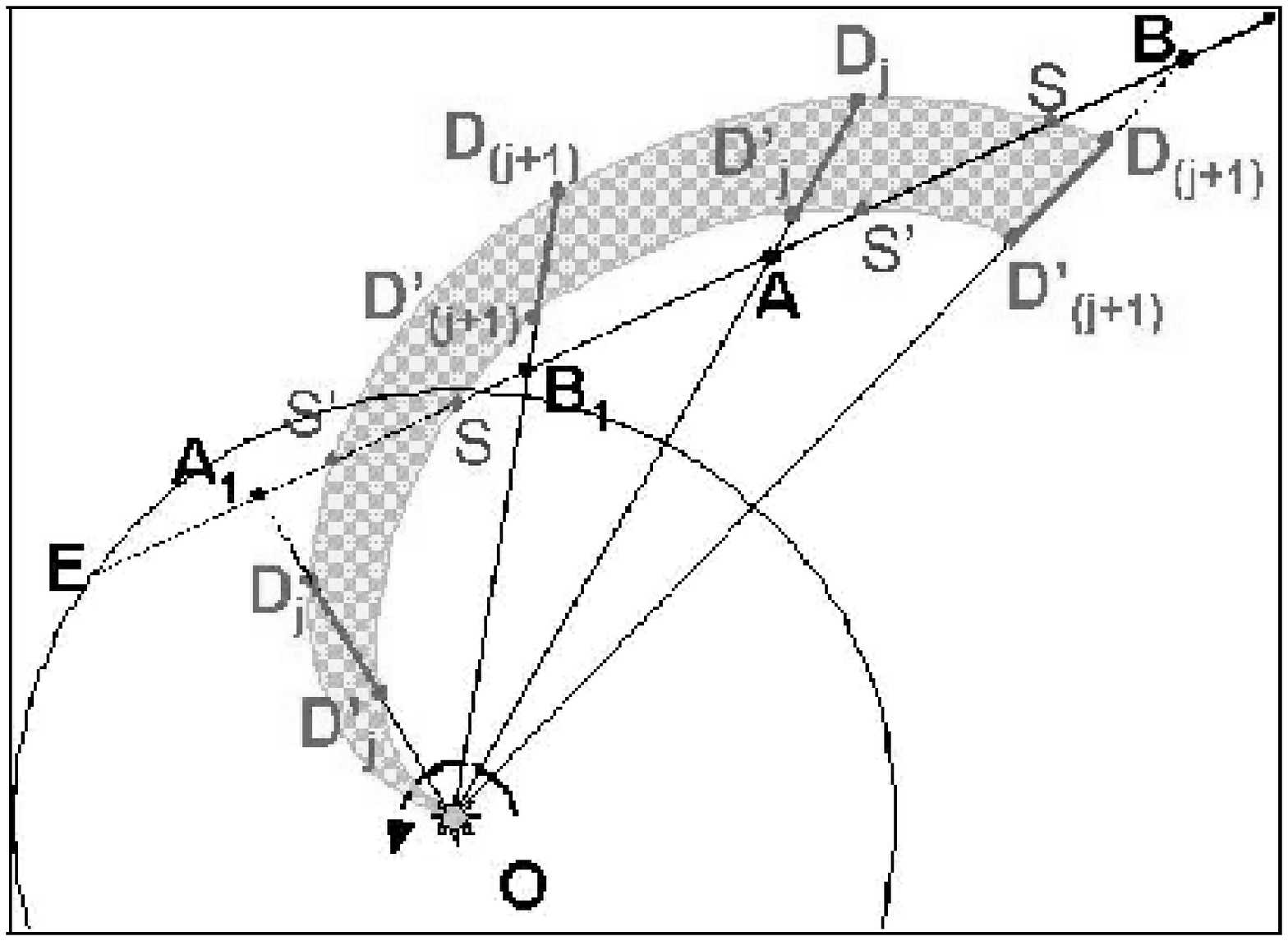}
}
\caption[\small{\textit{}}]
{\footnotesize{General geometries of the SW impact on the LOS sub-segment $SS'$ as seen from the North ecliptic pole. 
\textit{Panel a}: LOS pointing forward on the orbit ($AB = ds_j$, $OA = r_j$, $OB = r_{j+1}$).
\textit{Panel b}: LOS pointing backwards on the orbit ($AB = ds_j$, $OA = r_j$, $OB = r_{j+1}$). 
$A_1B_1 = ds_j$, $OA_1 = r_j$, $OB_1 = r_{j+1}$ have equivalent annotations, but $SS'$ is calculated differently 
for this case, since the spiral is approaching the observer (see details in the text).}}
\label{LOS_geo}
\end{figure*}

At each instant $T_i$ we define the form of the Parker spiral by calculating the radial distance 
$OD_j = D_j = V_{SW}\, (T_i - T_{d_j})$ (if $T_i \geq  T_{d_j}$, else $D_j = 0$) the front of the spiral has reached, 
and the total width $D_jD'_j = \Delta D_{tot} = \Delta t\, V_{SW} = D_{(j+1)}D'_{(j+1)}$ of the spiral. $T_{d_j}$ is the instant 
when the spiral leaves the Sun along the radial $r_j$, depending on the radial propagation of the SW and the solar rotation 
(considered as solid for simplicity with a 27-day period), and defined by the equation \ref{dep_time}:
\begin{equation}\label{dep_time}
\small{T_{d_j} =  T_o - (1 AU\,/\,V_{SW}) +   \left\{\, \begin{array}{l}
-27\, \varphi _j / 360 ,\, if\, LOS\, backwards\\
\\
+27\, \varphi _j / 360,\, if\, LOS\, forward  \\
\end{array}\right.}
\end{equation}
where $T_o$ is the instant on which the SW enhancement is measured 
in the solar instruments, and $\varphi _j = \lambda _{obs} - \lambda _j$ is the angle $E\nHat OA$ (see fig.A.\ref{losfront}) 
between the observer position (defined by the helioecliptic longitude $\lambda _{obs}$) and the point $r_j$ of the LOS 
(defined by the helioecliptic coordinates $\lambda _j$, $\beta_j$). We also define the distance $AD_j = \Delta D_j = D_j - r_j$, 
with respect to the radial position $r_j$ of the begining of the segment $ds_j$, which will be used later on. 

When the segment $AB = ds_j$ is only partially affected by the SW enhancement, we need to define a new enhancement 
factor $f_x\, =\, (f_{SW} - 1)\frac{SS'}{ds_j} + 1$ to apply on the entire 
segment $ds_j$ to account for the difference between the sub-segment really affected, $SS'$, and the total length $ds_j$ 
of the segment. An equivalent adjustment is to be applied in the abundances variation factor: 
$[A]_x\, =\, ([A]_{SW} - 1)\frac{SS'}{ds_j} + 1$. We will at first describe the procedure for the simplest case, 
when the LOS is ponting forward (fig. A.\ref{losfront}), and then the more complicated case of the LOS pointing backwards 
(fig. A.\ref{losback}). In the two figures the length of segment $ds_j$ is exagerated with respect to real dimensions 
for better legibility.

\subsection{LOS forward}
We calculate the sub-segment $SS'$ indirectly by defining the segments $AS'$ and $SB$ : $SS' = ds_j - AS' - SB$. We assume that the 
arcs $D_jD_{(j+1)}$ and $D'_jD'_{(j+1)}$ 
are small enough to be considered as straight lines. This is 
reasonable, for the segments $ds_j$ are narrow, especially near the observer (0 to 5 AU) where the emissivity is maximum 
and needs to be defined accurately. Then, from the opposed triangles ($S'BD'_{(j+1)}$ and $S'AD'_j$) and ($SBD_{(j+1)}$ and $SAD_j$) 
we can have respectively:

\begin{equation}\label{as}
\small{\begin{array}{l}AS' =  ds_j\, \left( 1 + (OD'_j\,BD'_{(j+1)})/(OD'_{(j+1)}\,AD'_j) \right) ^{-1}\\
\\
\, = \left\{\begin{array}{l}{ds_j\, \left( 1 + \frac{(D_j - \Delta D_{tot})\, \,(\mid \Delta D_{j+1}\mid + \Delta D_{tot})}
{(D_{j+1} - \Delta D_{tot})\, \,(\mid \Delta D_j\mid - \Delta D_{tot})}\, \right) ^{-1}},\, 
	\tiny{if\, \Delta D_j > \Delta D_{tot}}>0 \\
	\\
	0,\, \tiny{if\, \Delta D_j \leq 0}\\
\end{array}\right.
\end{array}}
\end{equation}

\begin{eqnarray}\label{sb}
\small{\begin{array}{l}SB = ds_j\, \left( 1 + (OD_{\,(j+1)}\, AD_j)/(OD_j\, BD_{(j+1)}) \right) ^{-1}\\
\\
\, = \left\{\,  \begin{array}{l}{ds_j\, \left( 1 + \frac{D_{j+1}\, \,\mid \Delta D_j\mid }{D_j\, \,\mid \Delta D_{j+1}\mid }\, 
	\right) ^{-1}},\, \tiny{if\, \Delta D_{j+1} < 0}\\
	\\
	0,\, \tiny{if\, \Delta D_{j+1} \geq 0}\\
\end{array}\right.
\end{array}}
\end{eqnarray}

\subsection{LOS backwards}
In the second case of the LOS pointing backwards, the combined effect of the solar rotation and of the SW radial 
propagation acts in such a way that the LOS is affected starting at an intermediate point, dividing the LOS in two parts: 
(i) one on which the spiral is moving away from the observer, and (ii) a second (usually smaller) on which the spiral is 
approaching the observer (fig. A.\ref{losback}). For the first part, where the spiral is moving away from the observer, 
equations \ref{as} and \ref{sb} remain unchanged. In the second 
part, where the spiral is approaching the observer, we have named the segment $ds_j = A_1B_1$ in figure A.\ref{losback} to avoid 
confusion. Respectively, we have named $OA_1 = r_j$, $OB_1 = r_{j+1}$, $A_1D_j = \Delta D_j = D_j - r_j$ 
and $SS' = ds_j - A_1S' - SB_1$. Thus, we calculate the segments $A_1S'$ and $SB_1$ from the opposed triangles ($S'A_1D_j$ and 
$S'B_1D_{(j+1)}$) and ($SB_1D'_{(j+1)}$ and $SA_1D'_j$) respectively:

\begin{equation}\label{a1s}
\small{\begin{array}{l}A_1S' = ds_j\, \left( 1 + (OD_j\, \,B_1D_{(j+1)})/(OD_{(j+1)}\, A_1D_j) \right) ^{-1}\\
\\
\, = \left\{\,  \begin{array}{l}{ds_j\, \left( 1 + \frac{D_j\, \,\mid \Delta D_{j+1}\mid}{D_{j+1}\, \,\mid \Delta D_j\mid}\, 
	\right) ^{-1}},\, 
	if\, \Delta D_j < 0 \\
	\\
	0,\, if\, \Delta D_j \geq 0\\
\end{array}\right.
\end{array}}
\end{equation}

\begin{eqnarray}\label{sb1}
\small{\begin{array}{l}SB_1 = ds_j\, \left( 1 + (OD'_{(j+1)}\, \,A_1D'_j)/(OD'_j\, \,B_1D'_{(j+1)}) \right) ^{-1}\\
\\
\, = \left\{\,  \begin{array}{l}{ds_j\, \left( 1 + \frac{(D_{j+1} - \Delta D_{tot})\, \,(\mid \Delta D_j\mid + \Delta D_{tot})}
{(D_j - \Delta D_{tot})\, \,(\mid \Delta D_{j+1}\mid - \Delta D_{tot})}\, \right) ^{-1}},\, 
	if\, \Delta D_{j+1} > \Delta D_{tot} > 0 \\
	\\
	0,\, if\, \Delta D_{j+1} \leq 0\\
\end{array}\right.
\end{array}}
\end{eqnarray}
\\

\begin{figure*}
\centering
\includegraphics[width=0.75\textwidth ,height=!]{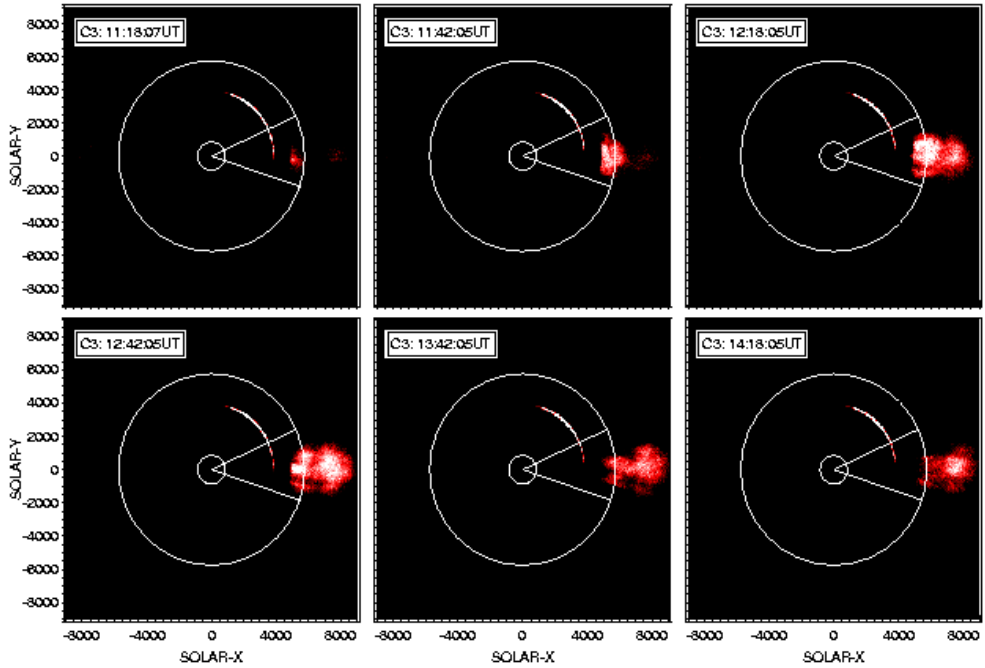}
\caption{\footnotesize{ON-LINE Material: LASCO C3 images of the 12/08/2000 10:35UT CME.The inner circle represents the Sun. 
The outer circle represents the sphere at 6.4 Solar Radii used to calculate the CME proton flux. The CME crosses this 
surface between 11:18UT and 14:18UT.}}
\label{lasc3}
\end{figure*}

\begin{figure*}
\centering
\includegraphics[width=0.75\textwidth ,height=!]{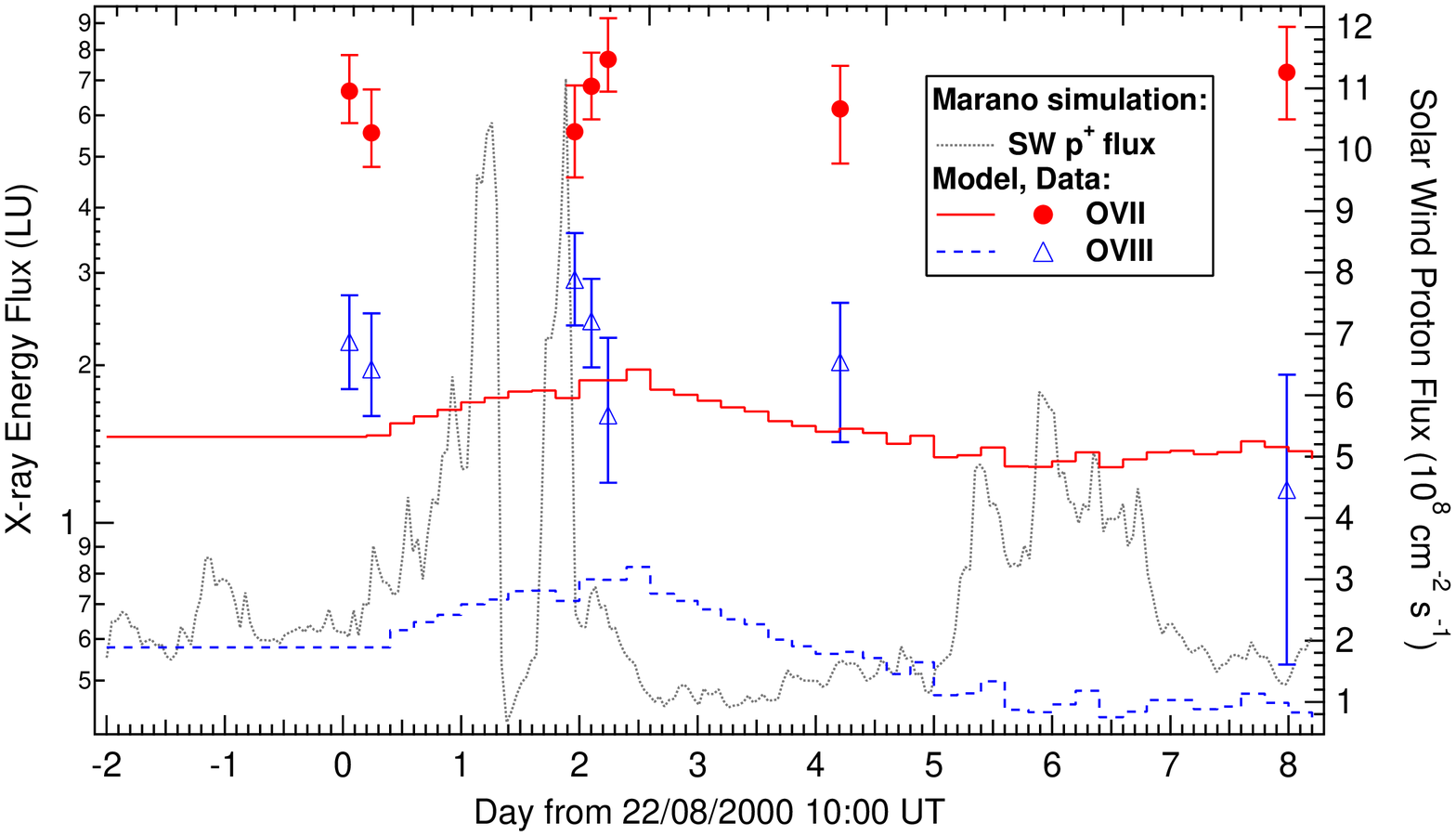}
\caption{\footnotesize{ON-LINE Material: Simulated lightcurves of O VII (red line) and O VIII (blue line) emission 
in LU for the Marano Field XMM exposures (22 - 27/08/2000 period). X-ray intensities are shown in logarithmic scale.
X-ray data are presented with red dots for O VII triplet and with blue triangles for O VIII line and positioned in 
mid-exposure time. SW proton flux is presented in the dotted curve.}}
\label{marano_lcurve}
\end{figure*}

\begin{figure*}
\centering
\includegraphics[width=0.75\textwidth ,height=!]{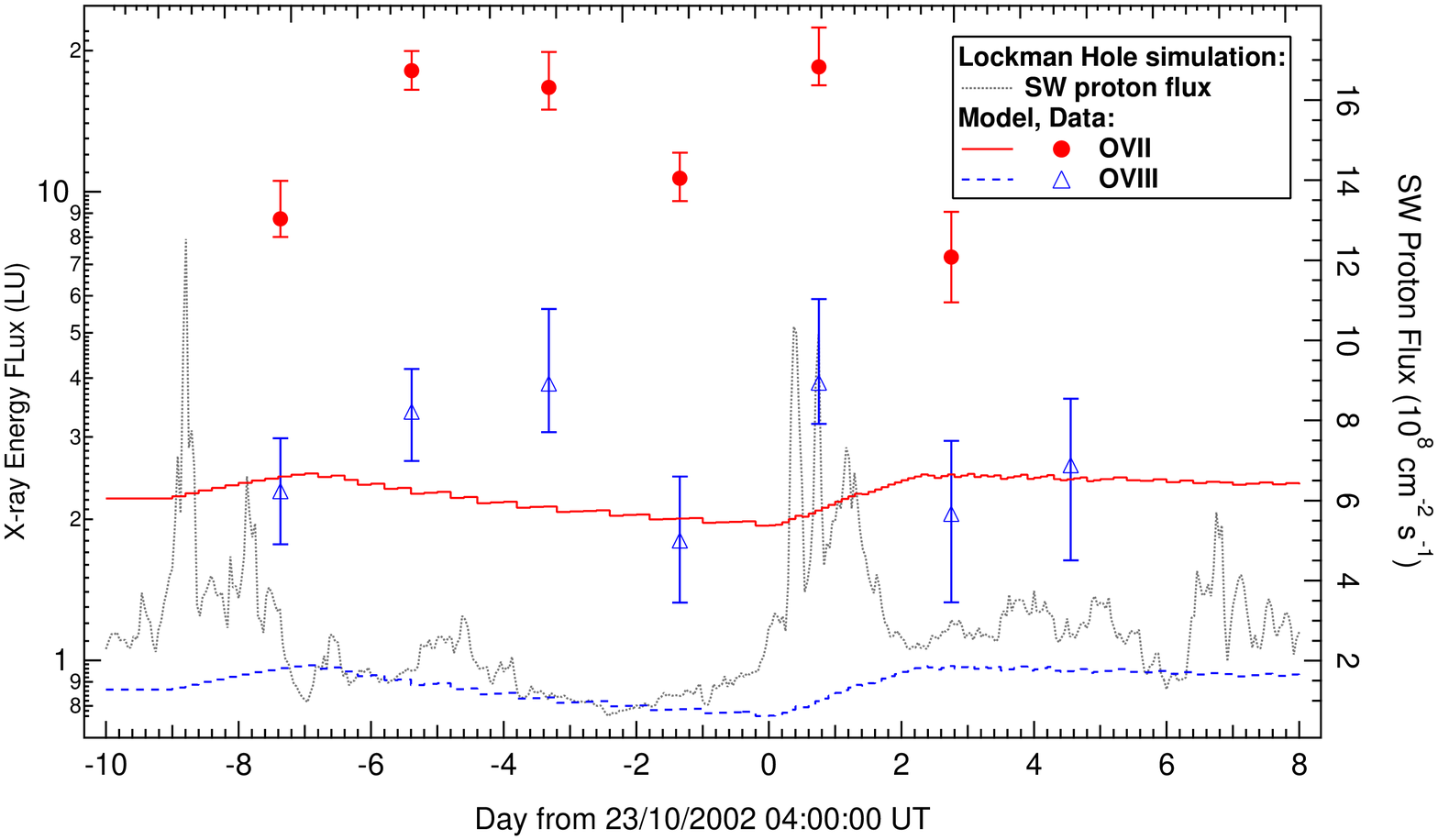}
\caption{\footnotesize{ON-LINE Material: Simulated lightcurves of O VII (red line) and O VIII (blue line) emission 
in LU for the Lockman Hole XMM exposures (15 - 28/10/2002 period). X-ray intensities are shown in logarithmic scale.
X-ray data are presented with red dots for O VII triplet and with blue triangles for O VIII line and positioned in 
mid-exposure time. SW proton flux is presented in the dotted curve.}}
\label{lock_lcurve}
\end{figure*}


\begin{thebibliography}{}

\bibitem[Andersson et al. 2002]{andersson02}
Andersson, B.-G., Idzi, R., Uomoto, A., Wannier, P. G., Chen, B., Jorgensen, A. M., 
2002, AJ, 124, 4, 2164

\bibitem[Arnaud 1996]{arnaud96}
Arnaud, K. A., 
1996, ASPC, 101, 17	
		
\bibitem[Chen et al. 1997]{chen97}
Chen, L.-W., Fabian, A. C., Gendreau, K. C., 
1997, MNRAS, 285, 3, 449
	
\bibitem[Cox (1998)]{cox98}%
Cox, D. P., 
1998, Lect. Notes Phys., 506, 121

\bibitem[Cravens et al. (1997)]{cravens97}%
Cravens, T. E., 
1997, GeoRL, 24, 1, 105	
	
\bibitem[Cravens (2000)]{cravens00}
Cravens, T. E., 
2000, ApJ, 532, 2, L153	
	
\bibitem[Cravens et al. (2001)]{cravens01} 
Cravens, T. E., Robertson, I. P., Snowden, S. L., 
2001, JGR, 106, A11, 24883

\bibitem[Edgar et al. 2006]{edgar06}
Edgar, R. J., Wargelin, B. J., Raymond, J. C., Slavin, J. D., Smith, R. K., Kharchenko, V., 
2006, HEAD meeting 9.



\bibitem[Freyberg 1994]{freyberg94}
Freyberg, M. J., 
Technische Univ. M$\ddot{u}$nchen, 1994.
	
\bibitem[Freyberg (1998)]{freyberg98}
Freyberg, M. J., 
1998, Astron. Nachr., 319, 1, 93

\bibitem[Freyberg et al. (2004)]{freyberg04}
Freyberg, M. J., Breitschwerdt, D., Alves, J., 
2004, Mem. Soc. Astron. Ital., 75, 509

\bibitem[Freyberg 2004]{freyberg04b}
Freyberg, M. J., 
2004, Ap\&SS, 289, 3, 229 
	
\bibitem[Fried et al. 1980]{fried80}
Fried, P. M., Nousek, J. A., Sanders, W. T., Kraushaar, W. L., 
1980, ApJ, 242, 987

\bibitem[Hartmann \& Burton (1997)]{hartmann97}
Hartmann, D., Burton, W.B., 
1997, Cambridge University Press.	
	
\bibitem[HSK07]{henley07xmm}
Henley, D. B., Shelton, R. L., Kuntz, K. D., 
2007, ApJ, 661, 1, 304. (HSK07)

\bibitem[Henley et al. (2007b)]{henley07suz}
Henley, D. B., Shelton, R. L., 
2007, to appear in the proceedings of ''The Extreme Universe in the Suzaku Era'', Kyoto, Japan, December 4-8, 2006.

\bibitem[Hobbs et al. 1986]{hobbs86}
Hobbs, L. M.; Blitz, L.; Magnani, L., 
1986, ApJ, 306, L109

\bibitem[Kappes et al. (2003)]{kappes03}
Kappes, M., Kerp, J., Richter, P., 
2003, A\& A, 405, 607	
	
\bibitem[Kharchenko \& Dalgarno 2000]{kharchenko00}
Kharchenko, V., Dalgarno, A., 
2000, JGR, 105, 18351
	
\bibitem[Kharchenko 2005]{kharchenko05}
Kharchenko, V.,  
2005, AIP Conf. Proc., vol. 774, 271

\bibitem[Koutroumpa et al. (2006)]{koutroumpa06}
Koutroumpa, D., Lallement, R., Kharchenko, V., Dalgarno, A., Pepino, R., Izmodenov, V., 
and Qu\'emerais, E., 
2006, A \& A, 460, 1, December II 2006, 289

\bibitem[Krumpe et al. (2007)]{krumpe07}
Krumpe, M., Lamer, G., Schwope, A. D., Wagner, S., Zamorani, G., Mignoli, M., Staubert, R., Wisotzki, L., Hasinger, G., 
2007, A\& A, 466, 1, 41	

\bibitem[Kuntz \& Snowden 2000]{kuntz00}
Kuntz, K. D., Snowden, S. L., 
2000, ApJ, 543, 1, 195	
	
\bibitem[Lallement et al. 2003]{lall03}
Lallement, R., Welsh, B. Y., Vergely, J. L., Crifo, F., Sfeir, D., 
2003, A\& A, 411, 447

\bibitem[Lallement (2004)]{lall04}
Lallement, R., 
2004, A\& A, 418, 143
	
\bibitem[Lisse et al. 1996]{lisse96}
Lisse, C. M., Dennerl, K., Englhauser, J., Harden, M., Marshall, F. E., Mumma, M. J., Petre, R., Pye, J. P., 
Ricketts, M. J., Schmitt, J., Trumper, J., West, R. G., 
1996, Sci, 274, 5285, 205	
	
\bibitem[Marano et al. (1988)]{marano88}
Marano, B., Zamorani, G., Zitelli, V., 
1988, MNRAS, 232, 111

\bibitem[McCammon \&\ Sanders 1990]{mccammon90}
McCammon, D, Sanders, W. T., 
1990, ARA\&A, 28, 657	

\bibitem[Morrison \& McCammon 1983]{morrison83}
Morrison, R., McCammon, D.,	
1983, ApJ, 270, 119
	
\bibitem[Pepino et al. 2004]{pepino04}
Pepino, R., Kharchenko, V., Dalgarno, A., Lallement, R., 
2004, ApJ, 617, 2, 1347
	
\bibitem[Raymond \& Smith (1977)]{raymond77}
Raymond, J. C., Smith, B. W., 
1977, ApJS, 35, 419

\bibitem[Robertson et al. (2001)]{robertson01}
Robertson, I. P., Cravens, T. E., Snowden, S., Linde, T., 
2001, Space Sci. Rev., 97, 1/4, 401	

\bibitem[Robertson \& Cravens (2003)a]{robertson03a}
Robertson, I. P., Cravens, T. E., 
2003a, GeoRL, 30, 8, 22 

\bibitem[Robertson \& Cravens 2003b]{robertson03b}
Robertson, I. P., Cravens, T. E., 
2003b, JGRA, 108, A10, pp. LIS 6-1
		
\bibitem[Sanders et al. 1977]{sanders77}
Sanders, W. T., Kraushaar, W. L., Nousek, J. A., Fried, P. M., 
1977, ApJ, 217, L87

	

\bibitem[Schwadron \& Cravens (2000)]{schwadron00}
Schwadron, N.A., and Cravens, T., 2000, ApJ, 544, 558

\bibitem[Shelton et al. 2007]{shelton07}
Shelton, R. L., Sallmen, S. M., Jenkins, E. B., 
2007, ApJ, 659, 1, 365

\bibitem[Snowden et al. 1990]{snowden90}
Snowden, S. L., Cox, D. P., McCammon, D., Sanders, W. T., 
1990, ApJ, 354, 211
	
\bibitem[SMV93]{SMV93}
Snowden, S. L., McCammon, D., Verter, F., 
1993, ApJ, 409, 1, L21 (SMV93)

\bibitem[Snowden 1993]{snowden93}
Snowden, S. L., 
1993, AdSpR, 13, 12, (12)103

	
\bibitem[Snowden \& Freyberg 1993]{snow_frey93}
Snowden, S. L., Freyberg, M. J., 
1993, ApJ, 404, 1, 403
	
\bibitem[Snowden et al. 1995]{snowden95}
Snowden, S. L., Freyberg, M. J., Plucinsky, P. P., Schmitt, J. H. M. M., Truemper, J., Voges, W., Edgar, R. J., 
McCammon, D., Sanders, W. T., 
1995, ApJ, 454, 643	
	
\bibitem[SCK04]{snowden04}
Snowden, S.L., Collier, M.RM, and Kuntz, K.D., 2004, ApJ, 610, 1182 (SCK04)

\bibitem[Smith et al. 2005]{smith05}
Smith, R. K., Edgar, R. J., Plucinsky, P. P., Wargelin, B. J., Freeman, P. E., Biller, B. A., 
2005, ApJ, 623, 1, 225

\bibitem[Smith et al. (2007)]{smith07}
Smith, R. K., Bautz, M. W., Edgar, R. J., Fujimoto, R., Hamaguchi, K., Hughes, J. P., Ishida, M., Kelley, R., Kilbourne, C. A., 
Kuntz, K. D., McCammon, D., Miller, E., Mitsuda, K., Mukai, K., Plucinsky, P. P., Porter, F. S., Snowden, S. L., Takei, Y., Terada, Y., 
Tsuboi, Y., Yamasaki, N. Y.,
2007, PASJ, 59, No.SP1, 141

\bibitem[Wang (1998)]{wang98}
Wang, Q. D., 
1998, LNP, 506, 503

\bibitem[Wargelin et al. 2004]{wargelin04}
Wargelin, B. J., Markevitch, M., Juda, M., Kharchenko, V., Edgar, R., Dalgarno, A., 
2004, ApJ, 607, 1, 596
	
\end{thebibliography}
\end{document}